\documentclass{emulateapj}
\shorttitle{UV-Bright Stellar Populations in M15}
\shortauthors{Haurberg et al.}
\usepackage{graphicx}
\usepackage{amssymb}
\begin{document}
\title{UV-Bright Stellar Populations and Their Evolutionary Implications in the Collapsed-Core Cluster M15}
\author{Nathalie C. Haurberg, Gabriel M. G. Lubell, Haldan N. Cohn, \\ and Phyllis M. Lugger}
\affil{Indiana University Department of Astronomy Bloomington, IN 47405, USA}
\email{nhauber@astro.indiana.edu, glubell@astro.indiana.edu, cohn@astro.indiana.edu, lugger@astro.indiana.edu}
\author{Jay Anderson}
\affil{Space Telescope Science Institute, Baltimore, MD 21218, USA}
\email{jayander@stsci.edu}
\author{Adrienne M. Cool}
\affil{Department of Physics and Astronomy, San Francisco State University, 1600 Holloway Avenue, San Francisco, CA 94132, USA}
\email{cool@stars.sfsu.edu}
\and
\author{Aldo M. Serenelli}
\affil{Max-Planck-Institut f\"{u}r Astrophysik, Karl-Schwarzschild-Str. 1, Garching, 85471, Germany}
\email{aldos@MPA-Garching.MPG.DE}

\begin{abstract}

We performed deep photometry of the central region of Galactic globular cluster M15 from archival Hubble Space Telescope data taken on the High Resolution Channel and Solar Blind Channel of the Advanced Camera for Surveys.   Our data set consists of images in far-UV (FUV$_{140}$; F140LP), near-UV (NUV$_{220}$; F220W), and blue (B$_{435}$; F435W) filters.  The addition of an optical filter complements previous UV work on M15 by providing an additional constraint on the UV-bright stellar populations.  Using color-magnitude diagrams (CMDs) we identified several populations that arise from non-canonical evolution including candidate blue stragglers, extreme horizontal branch stars, blue hook stars (BHks), cataclysmic variables (CVs), and helium-core white dwarfs (He WDs).  Due to preliminary identification of several He WD and BHk candidates, we add M15 as a cluster containing a He WD sequence and suggest it be included among clusters with a BHk population. 

We also investigated a subset of CV candidates that appear in the gap between the main sequence (MS) and WDs in FUV$_{140}-$NUV$_{220}$ but lie securely on the MS in NUV$_{220}-$B$_{435}$. These stars may represent a magnetic CV or detached WD-MS binary population. Additionally, we analyze our candidate He WDs using model cooling sequences to estimate their masses and ages and investigate the plausibility of thin vs. thick hydrogen envelopes.  Finally, we identify a class of UV-bright stars that lie between the horizontal branch and WD cooling sequences, a location not usually populated on cluster CMDs. We conclude these stars may be young, low-mass He WDs. 
\end{abstract}

\keywords{globular clusters: individual (M15) -- binaries: close -- stars: Population II -- white dwarfs -- stellar dynamics -- ultraviolet: stars}

\section{INTRODUCTION}
\label{sec:intro}

The Galactic globular cluster (GC) M15 is the archetypal core-collapsed GC with an extremely small and dense core \citep[as measured most recently by][]{vdb06} and thus has been an object of interest for many observers.  Though convincing models for core collapse have existed for some time \citep[e.g.,][]{cohn80}, many questions remain and recent attention has turned to how core collapse might affect the stellar populations in the central regions of GCs.

These high density central regions \citep[M15: $\rho_{0} \approx 7 \times 10^{6}$ M$_{\odot}$ pc$^{-3}$;][]{vdb06} provide an environment that can be ideal for the production of many types of exotic binaries. In such a setting, the encounter rate is sufficiently high that one expects to find an increased number of close binary systems due to dynamical interactions between stars that can harden primordial binaries and form new binary systems \citep{pool03}.  Because of the large amount of binding energy (sometimes comparable to the binding energy of the cluster) contained in close binary systems, interactions of these systems can have a significant effect on the dynamical evolution of the cluster \citep{hut92} and thus are of particular importance in understanding effects of core collapse.

Close binary systems, especially those containing a degenerate remnant, are expected to be overabundant in GC cores due to both the preferential formation of close binaries in high density environments as well as the effects of mass segregation. Many such systems are X-ray sources and therefore should be identified in X-ray surveys; M15 contains six X-ray sources with optical counterparts: two low-mass X-ray binaries (LMXBs) - AC211 \citep{aur84} and M15 X-2 \citep{whiang01, die05}, a dwarf nova cataclysmic variable (CV) - M15 CV1 \citep{sha04}, two suspected dwarf novae CVs whose optical counterparts are unclear \citep{han05}, and a quiescent LMXB - M15 X-3 \citep{hei09}.  In this work we probe into these close binary populations through the use of UV photometry and color-magnitude diagrams (CMDs).

Studies of M15's CMD have previously been used to examine many features of the cluster's stellar population. The blue straggler (BS) population was examined by \citet{yan94}, the horizontal branch (HB) and giant branches were analyzed by \citet{cholee07}, and UV studies by \citet{marpar94, marpar96} probed into the extremely blue objects, including white dwarfs (WDs). Most recently, \citet{die07} performed a UV analysis of M15, focusing on the variable stars in M15's core as well as performing a basic breakdown of the UV-bright stellar populations. This study revealed a substantial array of ``gap'' stars that appear in the region of the CMD surrounded by the HB, main sequence (MS), and WD cooling sequence. This gap zone had previously been dubbed the ``CV zone" by \citet{kni02} due to the likelihood of CVs appearing in this part of the CMD.

Optical CMDs of the central region of M15, though relatively deep, emphasize populations above the MS turn-off (MSTO). With perhaps a few exceptions, it is unlikely that CVs would be visible so far up on an optical CMD \citep[e.g.,][]{yan94, vdm02}. Observations in the UV, however, result in a distinct CMD morphology that highlights populations often not included on optical CMDs. In such diagrams, the gap zone is emphasized which can result in a larger list of potential CVs than that which optical filter combinations would yield. UV CMDs are also useful in identifying the extremely blue features of the HB and WD sequences in clusters.  M15 is a low-metallicity cluster \citep[{[}Fe/H{]} $= -2.26$;][]{har96}, which results in a correspondingly blue HB. In addition, M15 is known to possess an extended blue HB tail \citep{moe95} and, as we discuss in this paper, possibly harbors a small population of extreme-HB (EHB) stars which are hotter than the hottest HB stars predicted by canonical stellar evolution theories.  Furthermore, UV CMDs have the unique ability to photometrically distinguish a set of He-rich EHB stars known as ``blue hook (BHk) stars" which are characterized as being apparent EHB stars that are unusually faint in the UV \citep[e.g.,][]{whi98, dcr00, bro01}.

Another intriguing close-binary population expected to be found in the central regions of GCs are helium-core white dwarfs (He WDs).  He WDs are believed to arise in cases where mass loss on the red giant branch (RGB) is severe enough that the remaining mass in the H shell is too small to ignite core He-burning, resulting in an exposed He-core.  They are predicted to lie in sequences slightly brighter and redder than the canonical CO WD cooling sequences due to their lower mass and therefore larger radii \citep[e.g.,][]{benalt98}. The only cluster in which there is a confirmed sequence of He WDs is NGC 6397 \citep[][and references therein]{str09} though it has also been suggested that a significant fraction of the WDs in $\omega$ Cen are He WDs \citep{cal08}. A small number of He WDs are also known in each of several other nearby clusters as counterparts to ultracompact X-ray binaries \citep{and93, die05} and millisecond pulsars \citep{edm01, fer03, sig03}; in addition one was found in M4 by \citet{oto06} as a companion to a subdwarf B star and one was identified by \citet{kni08} in 47 Tuc using UV spectroscopy.

We performed PSF-fitting photometry on optical, near-, and far-UV images of M15's central region obtained by the Hubble Space Telescope (HST) using the Advanced Camera for Surveys (ACS).  This data set is a unique one: confined neither exclusively in the UV nor optical regimes; we analyzed M15 using colors that allow us to probe deeply into the optically faint yet UV-bright populations initially uncovered in M15 by \citet{die07}. By including an optical filter, we further investigate the photometric properties of these populations with an additional color and the more well-studied optical CMD morphology. In \S \ref{sec:obs} we describe the observations and photometry. In \S \ref{sec:cmd} we provide a breakdown of the various UV-bright populations. Section \ref{sec:analysis} contains our analysis, with \S \ref{sec:wdcool} containing the description our WD cooling models and analysis of the WD population, \S \ref{sec:raddist} describing our analysis of the radial distribution of populations believed to be of binary origin, and \S \ref{sec:discbhk}-\ref{sec:discbrightgap} focussing on the potential EHB, BHk, He WD, and CV populations, respectively.  Finally, we discuss the implications of our results in \S \ref{sec:discussion} and summarize our results in \S \ref{sec:conclusions}.

\section{OBSERVATIONS \& PHOTOMETRY}
\label{sec:obs}

Our data consist of HST archival images taken with ACS over several epochs in three filters.  The filters used were F435W, F220W, and F140LP; the first of these is similar to Johnson B, while the F220W and F140LP bands are in the near- and far-UV, respectively. The F435W data consisted of 13 separate frames, from GO-10401, taken on the High Resolution Channel (HRC) all having exposure lengths of 125 s.  The F220W and F140LP data, from GO-9792, were taken with the HRC and Solar Blind Channel (SBC), respectively.  The F220W data were taken in eight exposures of 290 s for a total exposure time of 2320 s. The F140LP data were taken in a manner designed to optimize detection of variability and consist of 90 individual exposures that result in a total exposure time of 24,800 s. These two data sets are described in detail by \citet{die07}.  Henceforth, the three filters will be referred to as B$_{435}$, NUV$_{220}$, and FUV$_{140}$. 

\begin{figure}[htbp]
\centering
\includegraphics[width=1\textwidth]{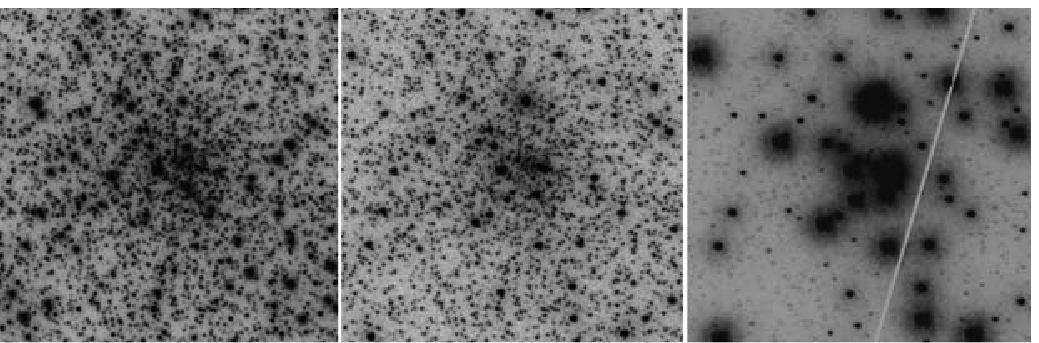}
\caption{The central 10\arcsec $\times$10\arcsec of the B$_{435}$, NUV$_{220}$, and FUV$_{140}$ images (from left to right). North is up in all 3 images.}
\label{fig:master_images}
\end{figure}

Despite the high resolution of the raw data, the individual frames were insufficient for resolving most stars in and around the cluster core.  Thus, to increase the resolution, the image data were combined using a computer program developed by one of us (J. A.).  The program is similar to drizzle with the pixfrac parameter set to 0, except that the transformations from each exposure to the master frame are based purely on the locations of the bright stars --- no image header information contributed.  The resolution of the resulting images was approximately doubled and was sampled with a pixel size of 0.0125$\arcsec \times$ 0.0125$\arcsec$ for all three filters. The central 10$\arcsec$ $\times$ 10$\arcsec$ of the stacked images are shown in Figure \ref{fig:master_images}; the complete field of view is 29$\arcsec$ $\times$ 26$\arcsec$ for the HRC and 39$\arcsec \times$ 31$\arcsec$ for the SBC.

We performed photometry using the software package DAOPHOT II \citep{ste90}. The detection threshold was determined by trial and error for each image individually in order to ensure that the majority of the sources were detected while still limiting the number of spurious detections.  The IRAF task \texttt{xyxymatch} was used to match individual stars across the filters.  A tolerance of 2 pixels was used to match between the B$_{435}$ and NUV$_{220}$ frames, while a 3-pixel tolerance was used between NUV$_{220}$ and FUV$_{140}$ frames. The difference in tolerances proved necessary on account of differing geometric distortions in fields from the HRC and SBC.  Suspected edge detections, spurious detections, and diffraction artifacts that passed through this routine were inspected by eye and rejected when necessary. 

Between 130 and 150 individual stars were selected from each image as models for a point-spread function (PSF).  Stars were chosen based on their visual appearance and the relative crowding of their neighborhoods. The PSF was determined in an iterative fashion with a careful effort to distinguish between PSF artifacts and close neighbors. The final quadratically-varying PSF was used with the ALLSTAR PSF-fitting photometry routine in order to determine the instrumental magnitudes.

Magnitudes were calibrated and transformed into the STMAG system in a similar manner to that outlined in detail in the HST ACS Data Handbook and by \citet{sir05}. Since ALLSTAR performs PSF fitting photometry, we first had to calculate an initial offset between the magnitudes from a small aperture, chosen to be 0.05$\arcsec$, and the magnitudes produced by ALLSTAR.  This calibration was performed with the same set of stars used to determine the PSF due to the relative brightness and isolated nature of these stars.  The offset was quite uniform across the field, with a standard deviation for the calibration stars not exceeding 0.016 magnitudes. Since the field is extremely crowded (see Fig. \ref{fig:master_images}) it was impossible to obtain a magnitude offset for the suggested 0.5$\arcsec$  aperture. Thus, in order to make the final calibrations for our magnitudes, we used the encircled energy fractions tabulated in \citet{sir05} and applied the STMAG zero point following the process outlined in the HST ACS Data Handbook. 

The calibrated data set was directly compared to the data set used in \citet{die07} (hereafter D07).  Of the 1913 stars with both NUV$_{220}$ and FUV$_{140}$ detections included in the data set used by D07 (A. Dieball, private communication 2007) we recovered a unique match for 1813.  In addition, we detected NUV$_{220}$ sources for 37 FUV$_{140}$ detections that had no matched NUV$_{220}$ source in the D07 data set. The 100 stars from D07 that we did not recover are not surprising; many of these stars are very dim and some lay in regions of the image that were not included in our master images (such as the occulting finger in the F140LP frame) because we relied solely on the combined frames for source detection.  In addition, some of the D07 FUV$_{140}$ sources appeared as two separate sources in our catalog, indicating that we possibly resolved some stars that appeared blended in their master images. Our final data set contained a total of 10,728 individual stars detected in at least two filters, necessarily including F220W.  This consisted of 3052 individual stars detected in both FUV$_{140}$ and NUV$_{220}$, 2943 of which were also detected in B$_{435}$.  Of the 1813 stars matched to sources from the D07 data set, 1777 were detected in B$_{435}$ as well. The discrepancy between our total number of FUV$_{140}$ sources, which totaled to 3961, and that of D07, which totaled to only 2137, seems to be due to our use of a lower threshold for source detection which resulted in the inclusion of on the order of 2000 faint sources that were not included in the D07 data set.  We expect that most of the false detections in this enlarged number of FUV$_{140}$ sources were filtered out by object matching across filters and our visual inspections of individual objects.

While most of our FUV$_{140}$ and NUV$_{220}$ photometry is consistent with the similar results presented in D07, we note that for many of the variable stars our FUV$_{140}$ magnitudes are systematically dim as compared to those of D07.  This is likely a consequence of having performed photometry on combined frames that had gone through a rejection routine designed to remove bad pixels, warm pixels, and other artifacts.  This routine rejected any pixel inconsistent at the 3 sigma level and therefore would affect any stars with variability that exceeded this threshold. 

\section{THE CMDs}
\label{sec:cmd}

Photometric results are illustrated in the CMDs presented in Figures 2-4.   In each figure we have identified various stellar populations based on their position in a given CMD (i.e. FUV$_{140}$-NUV$_{220}$ vs. FUV$_{140}$) and then plotted those same stars on a different CMD (i.e. NUV$_{220}$-B$_{435}$ vs. B$_{435}$) using a scheme that allows the stars, or groups of stars, to be tracked between the diagrams.  This technique allows us to better evaluate the photometric nature of a star or population by examining its position on multiple CMDs and additionally is useful for understanding the unfamiliar morphology of the UV CMDs.  

We have included multiple versions of the same CMD, with different identification schemes for the stellar populations, in order to clearly illustrate the photometric properties of each population we analyzed. Individual components of the various diagrams are discussed in the following sections. Unless otherwise stated, the differentiation between stellar populations was based on the overall appearance of the CMD and the location of the stars on the CMD.  We caution that this non-quantitative method will result in a small amount of confusion between the individual groups, but should not have a significant effect on our results. All magnitudes are given in the STMAG system.  

\subsection{Photometric results for M15 X-2, CV1, AC211 and H05X-B}
\label{sec:known}

We have obtained photometric measurements for 4 X-ray sources previously identified with close binary systems (labeled and plotted as  squares in Fig. 2-4).  These systems are include the confirmed dwarf nova (DN), M15 CV1 (hereafter CV1), identified by \citet{sha04} as well as optical counterparts to the two LMXBs: AC211, the optical counterpart to 4U 2127+119 as identified by \citet{aur84}, and M15 X-2 (hereafter X-2), identified as the optical counterpart of an ultracompact X-ray binary corresponding to X-ray source CXO J212958.1+121002 by \citet{whiang01} and \citet{die05}.  We also have provided photometry in NUV$_{220}$ and B$_{435}$ for a faint X-ray source identified in \citet{han05} as a possible DN or a quiescent soft-X-ray transient (qSXT; their ``source B'').  We refer to this source, hereafter, as H05X-B and discuss it in detail in \S \ref{sec:H05XB}.  H05X-B was too dim to be detected in FUV$_{140}$ (see Figure \ref{fig:H05X-B}), thus it is not included on CMDs which include this filter.  

Photometric data for these four objects are listed in Table \ref{tab:known}.  Our NUV$_{220}$ and FUV$_{140}$ magnitudes for X-2, AC211, and H05X-B are consistent with those of D07, within the errors.  We have also provided photometry in B, thus adding to the photometric information for these sources.

CV1 was not detected in the D07 FUV$_{140}$ master image as they found this star too faint and too near a bright neighbor to be detected.  But, they do detect the star during an outburst phase in one of the observing epochs, therefore confirming it as a DN.  In our FUV$_{140}$ master image, the source corresponding to CV1 is difficult to resolve by eye but the FIND task in DAOPHOT II was able to identify it as a discrete source (see Figure \ref{fig:CV1}).  However, when run through the PSF-fitting photometry of ALLSTAR, the proximity to the bright neighbor proved an issue and caused the star to be subtracted as part of the PSF of the neighboring star; thus we used a FUV$_{140}$ magnitude derived from aperture photometry. We report the NUV$_{220}$ and B$_{435}$ photometry of CV1 with equal confidence to stars of similar magnitudes, but note that since we were not able to use PSF-fitting for our FUV$_{140}$ photometry of CV1 the detection may be more questionable and our results less reliable.  

The NUV$_{220}$ magnitude for CV1 reported by D07 is significantly brighter than our reported magnitude.  As noted in the erratum to that paper, this is likely related to the fact that we performed PSF- fitting photometry while D07 used aperture photometry \citep{dieerat10}.  With aperture photometry, a  significant amount of flux from the bright neighbor star would likely end up in the aperture for the CV1 counterpart, and thus cause the calculated magnitude to be too bright.  When we performed aperture photometry with an aperture of comparable size, it resulted in a magnitude consistent with that in D07.

\subsection{The Main Sequence and Red Giant Branch}
\label{sec:MSandRGB}

The MS extends approximately 3.5 magnitudes below the turn-off in B$_{435}$ (see small orange dots in Figure \ref{fig:b_cmds}a).  The MS has a width of about 0.5 magnitudes just below the turn-off and widens somewhat at the faint end.  There is scatter to either side of the MS, some of which likely is due to photometric error, but may also indicate potential binary systems. The detection limit at B$_{435}$ $\simeq$ 22.5 and NUV$_{220}$$-$ B$_{435}$ $\simeq$ 2 is due to the depth of the NUV$_{220}$ data.  Since MS stars are relatively red, there are many fewer FUV$_{140}$ detections of MS stars and the MS in Figure \ref{fig:f_cmds}a extends only about one magnitude below the turn-off with a significant amount of scatter due its proximity to the detection limit. 

 The RGB (large red dots) is very well defined in Figure \ref{fig:b_cmds}a, forming a tight, narrow sequence stretching over 4.5 magnitudes in B$_{435}$.  The FUV$_{140}$ filter strongly suppresses the presence of the red giant population and the fact that a large number of RGB stars are detectable in the FUV$_{140}$ frame at all is a bit of a mystery.  It is most likely due to the ``red leak" phenomenon associated with the UV filters and discussed in \citet{chi07} and \cite{bof08}.  Red leak can greatly affect photometric measurements of red stars, but since the focus of this paper is UV-bright stars it should not affect our results.  Because it is very difficult to distinguish MSTO stars from RGB stars in the FUV$_{140} -$ NUV$_{220}$ vs. FUV$_{140}$ CMD we make the distinction only using the NUV$_{220} -$ B$_{435}$ vs. B$_{435}$ diagram.

\subsection{The Asymptotic Giant Branch}
\label{sec:AGB}

The asymptotic giant branch (AGB; small teal dots) is most easily identified in Figure \ref{fig:b_cmds}a as the slightly curved sequence turning away from the tip of the RGB and spanning about 2.5 magnitudes in color. The transition between the AGB and the HB is somewhat unclear; for the purposes of this paper we have defined the edge of the AGB just redward of the clump of variable stars belonging to the RR Lyrae instability strip.  Due to this loose distinction between the HB and AGB,  there may be some confusion between the blue edge of the AGB and the reddest HB stars, however this is not of particular concern as these stars are not the focus of this paper.

Like the RGB, the AGB is not as well defined in Figure \ref{fig:f_cmds}a because very little flux from these stars is emitted in the far-UV bandpasses.  However, the AGB can be identified as the sequence of stars connecting the tip of the RGB to the HB.  In Figures \ref{fig:b_cmds}b and \ref{fig:f_cmds}a there is a clear bend toward the blue at the point where the AGB transitions to the HB (FUV$_{140}$ $\approx$ 22.5).  

\subsection{The Horizontal Branch}
\label{sec:HB}
\subsubsection{Normal Horizontal Branch Stars}
\label{sec:normHB}

The main portion of the HB (plotted as large light blue dots on Fig. 2-4)  consists of a small set of red horizontal (RHB) stars and much larger set of blue horizontal branch (BHB) stars separated by variable stars in the RR Lyrae instability strip. A significant RHB population is not necessarily expected since metal-poor stars produce BHB stars; however, the presence of the RHB population in M15 is well documented \citep[e.g.,][]{buo85, pre06}. The BHB begins at the blue edge of the RR Lyrae strip extending toward the WD sequence.  For comparison we have plotted a theoretical zero-age horizontal branch (ZAHB; solid dark red and dark blue lines) generously provided by Santi Cassisi (private communication, 2009 \& 2010) and described in \citet{pie04}, on Figures \ref{fig:b_cmds} - \ref{fig:n_cmds}.  They were generated assuming a metallicity of Z$=$0.0001 (corresponding to an [Fe/H] $\approx -2.3$) and were adjusted for extinction following the method outlined in \citet{car89}.  We assumed a color excess of E(B$-$V)$=$0.1 \citep{har96} which resulted in a correction A$_{\lambda} =$ 0.81$\pm$0.11, 0.94$\pm$0.12, and 0.42$\pm$0.08 for FUV$_{140}$, NUV$_{220}$, and B$_{435}$, respectively .  

For the dark blue curve, a distance of 10.3$\pm$0.4 kpc \citep{vdb06} was assumed resulting in a distance modulus of 15.06.  This is in general the most accepted distance measurement for M15, however, it is clear in the CMDs including the FUV$_{140}-$NUV$_{220}$ color that using this distance modulus does not produce a good fit for the red side of the HB, as the ZAHB appears significantly brighter than the observed HB stars.  If the distance modulus is adjusted to 15.25, consistent with the distance modulus determined by \citet{kra03}, (solid dark red curve), a much better fit is achieved.  However, with the larger distance modulus the ZAHB is significantly dimmer than expected on the blue side of the HB.  So, neither distance modulus produces an ideal fit for the entire HB and we ascribe this to the limitations of these models in fitting our dataset.  Yet, since the fit is reasonable in both cases, we believe the models to be sufficient for the purposes of this paper.  The dotted lines associated with each curve represent the 1$\sigma$ error bars including both the error in distance and reddening.  For the dark blue curve, \citet{kra03} do not cite an error, so we assumed a reasonable error of 0.10 magnitudes in the distance modulus.

\subsubsection{Extreme Horizontal Branch Stars}
\label{sec:EHB}

As previously noted by many authors \citep[e.g.,][]{dor93, fer98, dal08} the hottest BHB stars owe their extremely blue colors to thin H envelopes which result in higher effective temperatures.  The thin envelope is believed to be due to significant mass loss on the RGB, likely to a close binary companion or possibly stellar winds. Because of the thin envelope these stars do not undergo the usual evolutionary progression to the tip of the AGB.  After leaving the HB, they have an insufficient mass in their hydrogen shell to enable the onset of the thermal pulsation phase; instead they either move off the AGB early and become post-early AGB (P-EAGB) or never ascend the AGB at all, becoming so called AGB-manqu\'{e} stars.  

This group, composed of the hottest BHB stars, is known as the extreme horizontal branch (EHB) and is often defined as the set of BHB stars with effective temperature greater than 20,000K  \citep[e.g.,][]{bro01, moe04, die09} but is more difficult to define observationally. Spectroscopically EHB stars correspond to subdwarf B (sdB) and subdwarf OB (sdOB) field stars \citep{heb87}, but photometric definitions differ between authors.  For clarity, we use the previously mentioned ZAHB models, and chose stars with T$\rm_{eff}$ $\geq$ 20,000K as EHB stars; using this method we identify 6 stars as EHB candidates (maroon pinched squares on Fig. 2-4). One of these candidates was not detected in B$_{435}$ due to its location near the edge of the B$_{435}$ image, so all 6 candidates only appear on Figures \ref{fig:b_cmds}b, \ref{fig:f_cmds}a, and \ref{fig:n_cmds}b and only Figure \ref{fig:f_cmds}a displays all 6 candidates with the expected shape and color.

\subsubsection{Blue Hook Stars}
\label{sec:bhk}

Five of the 6 potential EHB stars appear subluminous in the UV compared to BHB stars with similar temperatures (plotted as EHB stars in diamonds in Fig. 2-4). This indicates these as candidate blue hook (BHk) stars. (The subluminous nature of these 5 stars is perhaps most easily seen in Fig. \ref{fig:n_cmds}b.)  Defining EHB and BHk stars photometrically can be a very difficult task as there is no clear consistent photometric definition.  Here we have chosen to rigorously define EHB stars as only those with T$\rm_{eff}$ $\geq$ 20,000K and have chosen our BHk candidates as those that appear dimmer than the ZAHB models of similar temperature HB stars in the CMDs containing the FUV$_{140}-$NUV$_{220}$ color.  Clearly this definition depends somewhat on which ZAHB curves we use.  As can be seen in Figures \ref{fig:b_cmds}b, \ref{fig:f_cmds}a, and \ref{fig:n_cmds}b, four of the BHk candidates lie fairly clearly below the canonical ZAHB while the last one appears subluminous compared to the ZAHB using the smaller distance modulus (dark blue curve) but is more well within the error bars of the dimmer curve which uses larger distance modulus (dark red curve).  We still consider it a possible BHk star but recognize that its identification as such is weaker than the other four.

 Data on the stars we have identified as BHk candidates can be found in Table \ref{tab:uv}, which also contains photometric data for other UV bright sources whose classification is uncertain. These stars, as well as all the Table 2 stars, are plotted on Figure \ref{fig:cb}. 

 BHk stars are similar to EHB stars but have helium rich envelopes that cause increased opacity in the atmosphere and leads to their subluminous nature.  They are an intrinsically rare and relatively poorly understood population that have only recently begun to be identified \citep[e.g.,][]{bro01, bus07, sanhess08}.  At least two scenarios have been presented as to the origin of BHk stars.  One scenario, originally proposed by \citet{lee05}, is that BHk stars and the hottest HB stars may result from the standard evolution of a He-rich subpopulation in the cluster \citep[see also][]{die09, dancal08, moe07}.  The other scenario discussed by many authors \citep[e.g.,][]{bro01, dcr00, moe04} suggests that BHk stars lose a significant amount of mass along the RGB, likely to a close binary companion, such that they do not ignite helium at the tip of the RGB, but instead, undergo a helium-core flash as they begin to descend the white dwarf cooling sequence -- a so called late He-flash.

The ``blue hook" terminology is a reference to the appearance of this population on UV CMDs. While M15 does not currently have a recognized population of BHk stars, one helium-rich hot BHB star was spectroscopically identified by \citet{moe97} that would most likely appear as a BHk star on a CMD. This star is well beyond our field of view, so we can provide no further insight on it. However, the 5 candidates we have identified could establish the presence of a BHk population in M15, and may provide an important clue about the origin of these stars.  The significance of BHk stars in M15 is further discussed in \S \ref{sec:discbhk} \& \S \ref{sec:m15bhk}.
 
\subsection{White Dwarfs}
\label{sec:wd}

A prominent population of WD candidates that is in general agreement with the theoretical cooling models is apparent in the CMDs in Figures 2-4 (magenta pinched triangles). We have detected 64 WD candidates in FUV$_{140}$ and only 25 in B$_{435}$ (20 of which were also detected in FUV$_{140}$).  The majority of the WD candidates detected in FUV$_{140}$ were not detected in B$_{435}$ due to the faintness of these stars in B$_{435}$ and crowding problems that become particularly pronounced in the B$_{435}$ image.  As can be seen by examining the cooling curves plotted on Figures 2-4, some WD candidates appear to be more consistent with the cooling sequences for helium-core WDs (He WDs) than for normal carbon-oxygen WDs (CO WDs). This, along with the details of the WD population and models, is discussed in \S \ref{sec:wdcool}.

\subsubsection{Helium-Core White Dwarfs}
\label{sec:hewd}

He WDs are the remnant of a star that loses most of its hydrogen envelope on the RGB and therefore never undergoes a He-flash or subsequent HB and AGB evolution, but instead ends up cooling as a degenerate helium core with a thin hydrogen envelope.  He WDs are usually found as members of a binary system and are believed to arise primarily due to Roche lobe overflow and mass-transfer in a close binary system \citep[e.g.,][]{web75}. In dense environments such as GC cores it is possible that He WDs are formed by collisions involving RGB stars that result in a common envelope phase and eventual ejection of the envelope \citep{dav91}.  Although \citet{cascas93} show that in cases of extreme mass loss from winds it is possible for isolated stars to become He WDs, this seems an unlikely scenario for the production of a substantial number of He WD as the amount of mass loss required is much larger than can be explained by canonical stellar evolution alone. 

Based on theoretical studies, \citep[e.g.,][]{han03} GCs, especially those that have undergone core collapse, should possess observable He WD sequences.  However, only one GC (NGC 6397) has an extended He WD sequence currently identified \citep[e.g.,][]{cool98, edm99, tay01, han03, str09}. No He WD population has been previously identified in M15, but in \S \ref{sec:wdcool} we present strong evidence for the presence of a He WD sequence and further discuss this population in \S \ref{sec:dischewd}.

\subsection{Blue Stragglers}
\label{sec:bs}

The BS sequence appears as an extension of the MS to luminosities greater than the turn-off point.  It is generally believed that these stars are the products of a merger of two or more MS stars that, upon merging, produce a core hydrogen-burning star more massive than the MSTO.  There is an ongoing discussion of whether the most important formation mechanism for BSs is the gradual coalescence of a close binary systems or physical collisions between two stars \citep[e.g.,][]{fre04, map06, lei07, kni09}.  However, it is well agreed that the BSs found preferentially in the dense central regions of clusters.

We have identified 31 BS candidates from their position in Figure \ref{fig:b_cmds}a (small blue inverted triangles). However, it is difficult to judge where the MS ends and the BS sequence begins. It appears that Figure \ref{fig:f_cmds}a may be more insightful for distinguishing BSs from MSTO stars as the sequence of stars bluer than the MSTO is stretched out, thus making BSs more clearly distinct from the turn-off stars.  Using Figure \ref{fig:f_cmds}a we have identified approximately 53 BS candidates.

\subsection{Gap Objects}
\label{sec:gap}

We also have identified a significant set of stars that we term gap objects following D07 (plotted as bright green dots and open circles in Figures 2-4).  This population is composed of stars that populate the gap zone between the MS and WD regions of the CMD where there is an increased likelihood of finding CVs. The previously identified cataclysmic variable, CV1, appears in this region as expected.  

In Figure \ref{fig:f_cmds}a we identified 60 stars clearly populating the gap zone (thus classified as gap objects) however in Figure \ref{fig:b_cmds}a we identify only 22 gap objects, most of which lie near the MS or the faint-end photometric limits. Most of the sources identified as gap objects in Figure \ref{fig:f_cmds}a were detected in B$_{435}$ but rather than appearing in the gap region on NUV$_{220} -$ B$_{435}$ CMDs they appear primarily on the MS. This puzzling feature is discussed in detail in \S \ref{sec:discgap} .  We are confident in the significance of the gap population as identified in Figure \ref{fig:f_cmds}a, as these sources were all inspected by eye and D07 found a very similar population in their FUV$_{140}$ and NUV$_{220}$ photometry. 

\subsection{Bright Blue Gap Objects}
\label{sec:brightgap}

The population of stars that we identified as bright blue gap objects (pink asterisks in Fig. 2-4) consists of 8 stars (including X-2) detected in each filter that are bluer than the MS in both NUV$_{220} -$ B$_{435}$ and FUV$_{140} -$ NUV$_{220}$ by at least 1 magnitude but are found between the standard WD and HB sequences in brightness.  Canonical stellar evolution does not include stages expected to populate this region of the CMD for any significant time. WDs rapidly transiting from the post-AGB phase toward the WD cooling sequence may pass through this region, but the timescale is such that the likelihood of detecting even one star in such a phase is extremely small. Thus, from a canonical stellar evolution standpoint, the presence of several objects in this region is unexpected.  In Figures 2-4 it is however clear that the theoretical He WD cooling sequences run through this region and we will discuss the plausibility that these stars may be young He WDs in \S \ref{sec:brightgaphewd}.

This population is perhaps most clearly defined in Figure \ref{fig:f_cmds}a as the stars with with FUV$_{140} -$ NUV$_{220}<-0.5$ and 16.4 $\leq$ FUV$_{140}$ $\leq$ 18.4. Using this criteria we have identified 10 bright blue gap objects; however, in Figure \ref{fig:f_cmds}b, one appears more convincingly to be a hot WD and another appears on the RGB at NUV$_{220} -$ B$_{435} \approx$ 4.  The nature of the latter star is quite confusing; a plausible explanation is that we have detected a chance superposition of an RGB star and a very blue object, such as an HB star. When examined by eye, the center of light appears to be slightly inconsistent across the three images, supporting this hypothesis. 

It is unclear what the true nature of these bright blue gap objects is and we consider several possibilities in \S \ref{sec:discbrightgap}.  One of the 8 objects that appear as bright blue gap objects in both colors is X-2, raising the possibility that some of these stars could be close binaries with accretion disks. These stars may also be related to the BHk or He WD populations and each of these possibilities are discussed in \S \ref{sec:discbrightgap}.  Stars identified as bright blue gap objects are included in Table \ref{tab:uv} and Figure \ref{fig:cb} as UV8 and UV10-12.

\section {ANALYSIS}
\label{sec:analysis}

\subsection{White Dwarf Cooling Curves}
\label{sec:wdcool}

A set of cooling sequence models for both CO WDs (solid blue curves) and He WDs (dashed green curves and dotted purple curves) have been calculated and plotted on Figures \ref{fig:wd_thin} \& \ref{fig:wd_thick}. Figures 2-4 also include the CO WD cooling sequences (solid lines) and one set of the He WD cooling sequences (dashed lines) for orientation purposes. The code and input physics for these models are described in detail by \citet{ser02} and are calculated in the same manner as the models used in the analysis of the He WD population in NGC 6397 by \citet{str09}.  The models cover a mass range of M = 0.45 - 1.10 M$_{\odot}$ for the CO WDs and M = 0.175 - 0.45 M$_{\odot}$ for the He WDs (masses indicated in caption for Fig. \ref{fig:wd_thin} \& \ref{fig:wd_thick}). The models in Figures \ref{fig:wd_thin} \& \ref{fig:wd_thick} differ from each other in that the He WD models in Figure 8 (purple dotted lines) have thin hydrogen envelopes and the He WD models in Figure 9 (green dashed lines) have thick hydrogen envelopes. The thickness of the hydrogen envelope affects the color and cooling times for He WDs as discussed in \citet{ser02}, \citet{alt01}, and briefly in \S \ref{sec:dischewd} of this paper.  A distance of 10.3 kpc \citep{vdb06} was assumed and the cooling curves were adjusted for extinction in the same manner as the ZAHB.  We have included error bars in the upper portion of Figures \ref{fig:wd_thin} \& \ref{fig:wd_thick} that represent the 1$\sigma$ error in both distance and reddening for the cooling curves.

The He WD models have a progenitor metallicity of Z$=$0.0002 (corresponding to [Fe/H] $\approx$ $-2$), consistent with the derived metallicity for M15.  The CO WD models are all generated for progenitor stars with solar metallicity.  Since the evolution of CO WDs depends very little on nuclear burning and the cooling timescale is not metallicity dependent, we consider these tracks reasonably suitable for M15 despite the metallicity discrepancy.

From their location on the CMDs in Figures 2-4, we have identified a total of 73 stars that appear WD-like in at least one CMD. These stars are plotted as dots and open circles on Figures \ref{fig:wd_thin} \& \ref{fig:wd_thick}. It is somewhat difficult to distinguish CO and He WD candidates as there is large uncertainty in the photometry for the dim stars.  Table \ref{tab:wd} contains a summary of our assessment of these WD candidates as discussed in detail in the following paragraphs. 

Using their location on the CMDs, we determined 11 of these stars represent likely candidates for CO WDs, 45 appear to be likely He WD candidates, 10 appear to be good WD candidates but are ambiguous as to whether they belong to the CO or He WD population, and 2 are variable stars from D07 (plotted as filled triangles in Fig.  \ref{fig:wd_thin} \& \ref{fig:wd_thick}) that appear on the CO WD curves, but are more likely CV candidates.  The remaining 5 represent possible WD candidates but are subject to significant photometric scatter that makes their nature unclear. 

Of these 73 stars, only 20 were detected in all three frames (see filled black and red dots in Fig. \ref{fig:wd_thin} \& \ref{fig:wd_thick}). For 3 of these it is unclear whether they are truly WDs: they appear WD like in Figures \ref{fig:wd_thin}b \& \ref{fig:wd_thick}b but lie more than 0.5 magnitudes to the red side of the tracks in Figures \ref{fig:wd_thin}a \& \ref{fig:wd_thick}a, thus they may actually be CVs. This leaves us with 17 stars which we consider to be our strongest WD candidates because we have two-color data that allows us to better distinguish their true nature. 

We draw the following conclusions concerning these 17 stars: (1) Five are likely candidates for CO WDs.  Two are strong candidates that appear on the CO WD cooling curves in both diagrams and the other three are consistent with being either low-mass CO WDs or young, massive He WDs. (2) Seven are good candidates for He WDs because they appear reasonably consistent with the He WD cooling sequences in both diagrams and are clearly separated from the CO WD cooling curves. These are the 7 stars plotted as large red filled circles on Figures \ref{fig:wd_thin} \& \ref{fig:wd_thick}. (3) Five of the 17 stars are clearly WD-like but are ambiguous as to which population they belong since they appear on different cooling curves in each diagram.

We caution that, for many of these stars, the distinction between He WDs and CO WDs is very difficult and subject to assumptions such as the cluster reddening and distance (especially in CMDs using FUV$_{140}-$NUV$_{220}$ as these two quantities are the most sensitive to reddening).  The effect of the uncertainties in these quantities can be seen by examining the 1$\sigma$ error bars for the cooling curves which are plotted in the upper portion of Figures \ref{fig:wd_thin} \& \ref{fig:wd_thick}.  If we assume less reddening, some He WD candidates become better CO WD candidates, and if we assume more reddening, some CO WD candidates become He WD candidates.  Despite these uncertainties, we still find our claim that M15 possesses some population of He WDs to be strong.

In addition to the WD candidates discussed above, the bright blue gap objects are plotted on Figures \ref{fig:wd_thin} \& \ref{fig:wd_thick} (grey asterisks) because a subset of these objects appear to be consistent with the thick H envelope He WD cooling curves.  Of the 7 bright blue gap objects detected in all three frames (not including X-2), all appear to be consistent with being young, low-mass He WDs with thick hydrogen envelopes (see Fig. \ref{fig:wd_thick}).  The nature and origin of He WDs, the significance of the thin versus thick H envelope models, and the implications of the existence of He WDs in M15 is discussed in more detail in \S \ref{sec:dischewd}.

\subsection{Radial Distributions}
\label{sec:raddist}

Binary systems with a total mass larger than the average stellar mass in the cluster are expected to segregate towards the cluster core on the time scale of a half-mass relaxation time due to dynamical friction.  Since M15 is a core-collapsed cluster,  with a half-mass relaxation time of $t\rm_{r,hm}\approx1$ Gyr \citep{har96}, mass segregation should have taken place and any ``massive" binaries should be centrally concentrated. Many of the populations addressed in this paper (BS, BHk, EHB, He WD, CV)  are likely members of such binary systems, thus are expected to be segregated towards the cluster center. 
 
In Figure \ref{fig:raddist} we have plotted the cumulative radial distributions for the BSs, gap objects (which have been separated into two subsets, see \S \ref{sec:gapdist}), and bright blue gap objects, along with the radial distribution of HB stars for comparison. We chose the HB population as the reference population because it is bright enough that completeness issues should be minimized, even in the dense central regions.   To determine the statistical center of the cluster we used the MSTO population from Figure \ref{fig:b_cmds}a using the positions in the NUV$_{220}$ image.   We must note that any dim population will be drastically incomplete in the most central part of the cluster of the FUV$_{140}$ image because the extended PSF haloes and diffraction artifacts from several bright stars concentrated near the cluster center effectively mask many dim stars in this region (see Fig. \ref{fig:master_images}). 

 In order to analyze the significance of any apparent central concentrations, we performed Kologorov-Smirnov (KS) tests comparing the radial distribution of each population to our reference HB population.  The KS statistic gives the probability that the two samples being compared are consistent with being drawn from the same distribution; therefore, a lower KS probability corresponds to a more statistically significant difference between the distributions of the two samples, thus a more centrally concentrated sample.  

\subsubsection{Distribution of Bright Blue Gap Objects}
\label{sec:brightbluedist}

We find a KS probability of 0.04\% and 0.4\% for the bright blue gap objects (as identified in Figure \ref{fig:b_cmds}a \& \ref{fig:f_cmds}a, respectively) as being consistent with the HB sample.  This is a strong indication that these stars are more massive than the average HB mass in the cluster, likely due to a binary nature.

\subsubsection{Distribution of Blue Stragglers}
\label{sec:bsdist}

\begin{figure}[htbp]
\centering
\includegraphics[width=1\textwidth]{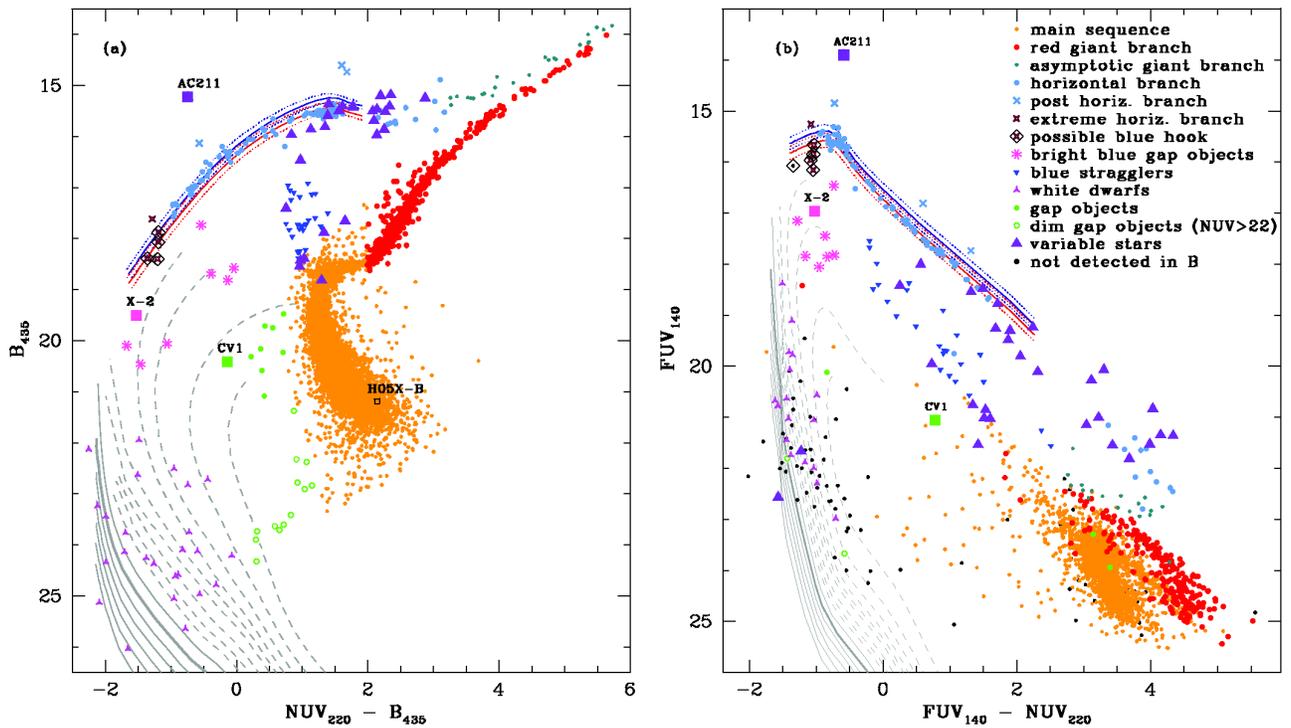}
\caption{Comparison of NUV$_{220} -$ B$_{435}$ and FUV$_{140} -$ NUV$_{220}$ CMDs.  Stellar populations are distinguished and color-coded according to their positions in 2a, the NUV$_{220} -$ B$_{435}$  CMD. See key upper righthand corner of 2b for symbol explanation. Theoretical WD cooling sequences for CO WDs (solid lines) ranging in mass from 0.45 - 1.10 M$_{\odot}$ and thick H envelope He WDs (dashed lines) ranging in mass from 0.175 - 0.45 M$_{\odot}$, as well as two ZAHBs (solid dark red line and dark blue lines) are included; see \S \ref{sec:wdcool} and \ref{sec:HB} for details.  All variable stars were identified as such by D07. The ``dim gap objects" (NUV $>$ 22) have been plotted as a separate group so they can be distinguished on all CMDS, the reason for this differentiation is described in \S \ref{sec:gapdist}. Squares mark optical counterparts of four X-ray sources discussed in the text.}
\label{fig:b_cmds}
\end{figure}

We find a KS probability of 1\% that the BS population, as identified from Figure \ref{fig:f_cmds}a, is consistent with the HB sample.  This indicates a significant segregation towards the cluster center. Yet, for the BSs identified from Figure \ref{fig:b_cmds}a, the KS probability is 30\%.  This is a surprising result considering that the latter set represents the brightest and presumably most massive BSs, which are expected to display the strongest effects of mass segregation.  Also, nearly every BS in Figure \ref{fig:b_cmds}a also was chosen as a BS in Figure \ref{fig:f_cmds}a, implying that it is the dimmer, less massive BSs that show the strongest degree of segregation; upon further inspection we find this to be true. 

If we consider the stars identified as BSs in Figure \ref{fig:f_cmds}a that were detected in all three frames and divide them into a ``bright" group with B$_{435}$ $<$ 18 and a ``dim" group with B$_{435}$ $\geq$ 18, we find that the ``bright BSs" have a KS probability of 13\% while the ``dim BSs" have a KS probability of 1\%.  This would seem to indicate that these dim BSs are significantly more centrally concentrated than the bright BSs.  Although a rigorous explanation for this phenomenon is beyond the scope of this paper, we suggest it may be a statistical issue due to the smaller number of BSs in Figure \ref{fig:b_cmds} (only 31 BSs selected from Fig. \ref{fig:b_cmds} vs. 53 selected from Fig. \ref{fig:f_cmds}).

Recently, a significant number of globular clusters have been discovered to have a bimodal BS distribution with a peak inside the core radius then a quick drop off in a region termed the ``zone of avoidance" followed by a second peak at several core radii \citep[e.g.,][]{fer93, map06, dal08}.  We don't expect to see such a distribution in our data since the expected radius for the drop off in M15 is well beyond our field of view. Following the method outlined in \citet{map06} with reasonable parameters for M15 we estimate a zone of avoidance at $r \approx 3.3 - 7.5\arcmin$ (10-20 pc), depending on the exact parameters used in the calculation.  Even though this is a relatively large range it is undoubtably beyond our field of view and near the half-mass radius of the cluster.

\subsubsection{Distribution of Gap Objects}
\label{sec:gapdist}

For the gap objects we find a KS probability of 89\% (from the Figure \ref{fig:b_cmds}a group) and 80\% (from the Figure \ref{fig:f_cmds}a group) as being consistent with the HB. This indicates that the gap objects do not show a central concentration.  But, since many of the gap objects are faint, this population suffers serious incompleteness issues, especially in the innermost regions of the cluster.  In an attempt to lessen this bias we have plotted the distribution of only those gap objects with NUV$_{220}$ $\leq$ 22.  The gap objects are plotted as two subsets on Figures 2-4 (NUV$_{220}$ $\leq$ 22: large bright green dots; NUV$_{220}$ $>$ 22: bright green open circles), and thus it can be seen that by using this criteria alone we were able to select a group that should have completeness levels that are more consistent with the other populations we are analyzing.

For this brighter subset (NUV$_{220}$ $\leq$ 22) we find a KS probability of 8\% and  64\% from the Figure \ref{fig:b_cmds} and \ref{fig:f_cmds}, respectively.  These are still relatively high KS probabilities and therefore do not necessarily indicate the presence of a statistically significant central concentration for these populations, but they do suggest that the brighter set of the population derived from Figure \ref{fig:b_cmds}a may be centrally concentrated to some degree.  Despite the moderate KS probabilities we report, we still consider it likely that at least some of these objects are close binary systems based on their photometric properties and suggest that the KS probabilities given here be regarded as upper limits due the completeness and crowding issues discussed above.  

\subsubsection{Distribution of Other Populations}
\label{sec:distother}

Both the EHB and He WD populations are possibly members of binary systems more massive than the average stellar mass in the cluster and therefore may be segregated to the center as well.  We, however, do not report on the radial distribution statistics for these populations due to issues of small number statistics and completeness, respectively.  The EHB is composed of only 6 stars, so the results from any statistical analysis we perform would not be robust.  The He WD population has enough candidate members to allow for more robust statistics, but the dim nature of these stars makes them suffer drastically from the completeness and crowding issues in the cluster center that have been discussed throughout this paper.  Therefore, there is an apparent dearth of these stars in the innermost region of the cluster which is almost certainly due to crowding issues that mask these dim stars.  However, because of this feature, our data is not sufficient to provide meaningful statistics on their radial distribution.   

\subsubsection{Comparison with D07}
\label{sec:compdist}

D07 also performed an analysis of the radial distribution of the BSs, and gap objects compared with the HB stars.  Consistent with our results, they found that the BSs are significantly more centrally concentrated than the HB.  However they find a KS probability of the two samples being from the same parent distribution of 14\%, which is much larger than the KS probability that we report of 1\%.  This is most likely due to differences between the stars classified as BSs; D07 classified 69 stars as BSs while we only classified 53 stars as such.  

Our finding that the gap objects do not necessarily appear to be centrally concentrated seems at odds with the finding in D07 that the gap population is the most centrally concentrated with a KS probability of only 8\% when compared with the HB stars.  However, D07 only considered stars with NUV$_{220}$ $<$ 21, so their result should only be compared with our result for the ``bright" gap objects with NUV$_{220}$ $<$ 22.  While the samples are still not entirely similar, it is striking that we find a much larger KS probability of 64\%, even for this brighter sample.  This could be, in part, due to the fact that we extend our cut a magnitude deeper and therefore include more stars and may have more issues concerning completeness near the cluster center.  However, a large contribution to this difference is almost certainly due to the fact that the gap population in D07 included the stars we have termed bright blue gap objects which are very strongly centrally concentrated.

\subsection{Blue Hook Candidates}
\label{sec:discbhk}

Of the many UV-bright stellar exotica that have been discussed thus far, evidence of a BHk population is of particular interest as it may provide an important clue about how these stars originate.  The presence of BHk stars in M15 is somewhat contrary to the hypothesis that BHk stars are the product of a He-rich subpopulation, because with a mass of only $4.4 \times 10^{5}$ M$_{\odot}$ \citep{vdb06} it is unclear whether M15 has a potential well deep enough to maintain sufficient gas from stellar ejecta to form a second generation of stars.  In addition, M15 does not show a split MS as expected for a cluster containing a second generation (e.g., NGC2808: Piotto et al. 2007; $\omega$ Cen: Bedin et al. 2004). \citet{dancal08} argue that the period distribution of RR Lyrae stars in M15 is best explained by the existence of a He-rich subpopulation, but note that their models produce HR diagrams which are not particularly good fits for the morphology of M15's HB. So, while a subpopulation in M15 should not be ruled out, it seems a somewhat unlikely explanation for BHk stars in this cluster.

Furthermore, the high central density of M15, $7 \times 10^{6}$ M$_{\odot}$ pc$^{-3}$ \citep{vdb06}, results in a very a high interaction rate weighing in favor of the late He-flash origin.  In this picture the He-flash occurs on the WD cooling sequence, thus it is necessary that there be enough mass loss during RGB phases to avoid the He-flash at the tip of the RGB . The high central density and high interaction rate of M15 should lead to an increased number of mass-transfer binaries and collisions that may strip the outer layers of an RGB star, thus increasing the potential for BHk stars to be formed this way.

\subsection{Helium White Dwarfs}
\label{sec:dischewd} 
  
We have identified a substantial number of He WD candidates (see \S \ref{sec:wdcool}).  However many of our candidates lie very near our detection limit and therefore suffer from significant photometric scatter and are more susceptible to being false detections.  For this reason, we will focus our analysis on those 7 He WD candidates that were detected in all three filters and appear as good candidates for He WDs in both Figures \ref{fig:wd_thin} \& \ref{fig:wd_thick} (plotted as large red filled circles).
 
 \subsubsection{Thick vs. Thin H Envelopes}
 \label{sec:thickthin}
One question of current interest about He WDs is how massive their hydrogen envelope is. The candidates identified here, in conjunction with the models discussed in \S \ref{sec:wdcool}, may provide some insight on this parameter.  The mass, and thus thickness, of the hydrogen envelope greatly affects the cooling timescale and, to a lesser extent, the photometric properties of a He WD of a given mass.  Since it is unclear what envelope mass is expected, we have included models for both ``thick" and ``thin" envelopes.  In the thick H envelope case nuclear burning (pp-chain) at the base of the envelope dominates the radiation and cooling time, while in the thin H envelope case the contribution from nuclear burning is negligible so the remnant cools via thermal radiation.  The models presented here represent the two opposite extremes in the the range of hydrogen envelope mass and therefore, as pointed out by \citet{str09}, should ``bracket reality." For further details of the models and the implications for ``thick" vs. ``thin" envelopes the reader is referred to \citet{ser02} and \citet{str09}.

\subsubsection{Photometric Masses}
\label{sec:photmass}

It is difficult to derive exact photometric masses from our candidate WD populations, however in Figure \ref{fig:wd_thick} the 7 strong He WD candidates seem to be well bracketed by the curves spanning the mass range from M $\approx 0.200$ M$_{\odot}$-- 0.275 M$_{\odot}$.  The 7 bright blue gap objects (grey asterisks) that appear to be He WD candidates in Figure \ref{fig:wd_thick} also appear to be most consistent with the models in this mass range.  The photometric mass range is much less well constrained in Figure \ref{fig:wd_thin}, as the set of 7 strong He WD candidates seem to span the mass range from M $\approx$ 0.175 M$_{\odot}$-- 0.350 M$_{\odot}$.  

It is nearly impossible to approximate a photometric mass using the entire set of 45 He WD candidates.  The candidates span the entire range of model He WD masses in Figure \ref{fig:wd_thick} and approximately 10 of the candidates appear significantly redder than the reddest model in Figure \ref{fig:wd_thin}.  It is possible that the reddest stars are not truly He WDs, but instead are either CVs or detached WD-MS binary systems that just happen to lie near the WD cooling sequence.  However, since this region is relatively well populated and reasonably well separated from the gap object population in color, we still present them as more likely WD candidates than gap objects. This still does not eliminate the possibility that some of our candidates may be CVs, detached WD-MS binaries, or even LMXBs especially as this region contains two variable stars classified as likely CVs based on the variability study by D07.

\subsubsection{Cooling Ages and Implication for Formation}
\label{sec:coolage}

Following the analysis in \citet{str09} we analyze the apparent cooling ages of the He WDs in order to gain insight into the possible formation mechanisms as well as investigate the plausibility of the thin vs. thick H envelope models. Focussing on the 7 strongest He WD candidates in the thin H envelope case (Figure \ref{fig:wd_thin}), we see that all 7 stars have cooling age $<$100 Myr, with the dimmest having a cooling age approximately equal to 100 Myr.  Therefore we can calculate an average formation rate of about 70 Gyr$^{-1}$.  Yet 4 of these stars have cooling ages of $<$ 2 Myr which implies an implausible formation rate in recent epochs of 2000 Gyr$^{-1}$.

We can estimate the rate at which stars are turning off the MS in M15 for comparison to the formation rate implied by the cooling ages of our WD candidates.  Adding our sample of MSTO stars that are clearly redward of the turn off point (NUV$_{220} -$ B$_{435} \approx$ 1.5) to our number of RGB stars, we have a total of about 800 stars in these phases.  Given that the time that a star spends in such phases is approximately 1.5 Gyr \citep{pols98}, we obtain a rate for stars turning off the MS of 530 Gyr$^{-1}$. This means that the formation rate implied by the 4 stars with cooling ages of $<$ 2 Myr using the thin H envelope models is significantly larger than the rate at which stars are leaving the MS and for this reason alone, we find that the formation rate implied from the thin H envelopes models to be unreasonable.  It is possible that these brighter candidates are not He WDs but are instead WD-MS detached binaries or CVs, so we can not rule out the thin H envelope model altogether with just this evidence.

Additionally, we should note that the estimation of the rate at which stars are turning off the MS can be calculated in several ways and may be sensitive to issues that are difficult to account for, such as mass segregation.  Another way to calculate this rate is to use the HB;  using the total number of stars found in the HB ($\approx$ 130) and an estimate for the lifetime of a HB star \citep[$\approx$ 100 Myr;][]{cas04} we get a turnoff rate of approximately 1300 stars Gyr$^{-1}$.  This is about 2.5 times larger than the rate obtained using the MSTO stars.  The reason that these two rates do not agree more closely is not clear but may be related to the uncertainties concerning the lifetime of post-MS phases such as the RGB and HB, as well as dynamical considerations such as mass segregation and stellar interactions.  However, even using this larger estimate for the turnoff rate in M15, we still find the formation rates implied by the thin H envelope models to be larger than the rate at which stars are turning off the MS and therefore unreasonable.

Using Figure \ref{fig:wd_thin} to evaluate the \textit{entire} population of He WD candidates, 38 of these appear to have cooling ages $\lesssim$ 100 Myr, with the remaining 7 either having a longer cooling age or being significantly redder than the most low-mass cooling sequence.  This implies a fairly extreme formation rate of 380 Gyr$^{-1}$ and would imply that between 25\% to 70\% of the stars turning off the MS are being formed into He WDs.  But, many of our candidates are not detected in all three filters so some of these candidate are likely to be falsely identified, and we therefore can not rule out the thin H envelope models on this evidence alone either. However, because we find significant doubt as to the feasibility of the formation rates implied by the thin H envelope models and the photometric mass range for these models seems much less well-constrained (see previous section) we will focus on the thick H envelope models for the the remainder of this section as they appear to be the more plausible models. 

For thick H envelope models (Figure \ref{fig:wd_thick}) the 7 strongest candidates appear to have cooling ages down to $\lesssim$ 1 Gyr. This implies a very reasonable average formation rate of 7 Gyr$^{-1}$, and if we include the 7 bright blue gap objects that also seem consistent with the thick H envelope models, we still find a reasonable average formation rate of 14 Gyr$^{-1}$.  Nine of these appear to have cooling ages $\lesssim$ 250 Myr implying a larger, yet not unreasonable, formation rate of 36 Gyr$^{-1}$.  In fact, this number is in excellent agreement with estimates calculated using our entire sample of He WD candidates in Figure \ref{fig:wd_thick} (see below).  In total, 30 of the candidates plus the 7 bright blue gap objects, appear to have cooling ages of $\lesssim$ 1 Gyr, implying a formation rate of 30--37 Gyr$^{-1}$.  The remaining 15 candidates appear to have cooling ages of $\gtrsim$ 1.5 Gyr, which will result in lower average formation rate in epochs more than 1 Gyr ago. Although, since this cooling age is well below the detection limit for many of the models we likely have only detected a fraction of the objects with cooling ages between 1 Gyr and 1.5 Gyr.  In summary, using the thick H envelope models we have obtained a formation rate between 7 Gyr$^{-1}$ and 37 Gyr$^{-1}$ depending on which potential He WDs are included in the sample.  

We can compare theses estimated formation rates with average collisional rate for RGB stars in the central portions of M15. We will use the formulation in \citet{bintre87} that 
\begin{displaymath}
\frac{1}{t\rm_{coll}} = 16\sqrt{\pi} n\sigma R_{*}^{2}(1+\frac{GM_{*}}{2R_{*}\sigma^{2}}).
\end{displaymath}
For our estimates we will confine the ``central portion" to the the central 0.1\arcmin\ (0.3 pc).  For the average stellar density, $n$, in this region, we will use the central luminosity density from \citet{har96} of $2.4 \times 10^{5}$ L$_{\odot}$ pc$^{-3}$ and an average M/L $\approx$ 2 for the central portion of M15 from \citet{vdb06} to get a density of $4.8 \times 10^{5}$ stars pc$^{-3}$ (assuming an average stellar mass for the central region of $\approx$ 1 M$_{\odot}$). Using this with the velocity dispersion for M15, $\sigma = 11$ km s$^{-1}$ \citep{geb94, dull97, dull03}, and the radius and mass of a red giant, R$_{*}$ = 10 R$_{\odot}$ and M$_{*}$ = 0.8 M$_{\odot}$, the collision rate for a given red giant is approximately 1 collision per  2 Gyr.  We found about 200 RGB stars in our data that lie within 0.1\arcmin\ of the cluster center, and thus predict 100 collisions involving an RGB star per Gyr in the central regions of M15.  

This predicted collision rate is high enough to account for the upper limits on the formation rates of He WDs calculated above (for the thick H envelope models) under the assumption that He WDs are in fact formed through collisions involving RGB stars.  However, we caution that our calculations here are very approximate and sensitive to our assumptions.

Mass loss in a close binary system is another possible mechanism for the formation of He WDs. So, again following \citet{str09}, we will assume that He WDs are formed primarily from primordial binaries and then estimate a lower limit on the binary fraction in the core of M15 by considering the formation rate of He WDs.  

 Our estimated range for the He WD formation rate of 7--37 Gyr$^{-1}$, leads to an implied binary fraction from 1\% to 7\% (using the MSTO to estimate the turnoff rate) or 0.5\% to 3\% (using the HB to estimate the turnoff rate).   These are in reasonable agreement with the results of \citet{geb97} who find a binary fraction of 7\% in M15.  However, we are considering only the fraction of binaries that produce He WDs, which itself is only a fraction of the binaries in the cluster, thus our estimate should be a lower limit on the binary fraction.  But, as shown previously in this section, collisions likely contribute to the formation of He WDs as well, and therefore we conclude that collisions involving RGB stars as well as mass transfer in close binary systems are both possible formation channels for He WDs in M15.
 
 \subsubsection{Bright Blue Gap Objects as He WDs}
\label{sec:brightgaphewd}

Many of the bright blue gap objects are consistent with the thick H envelope He WD cooling models (see Fig. 9).  This, combined with the very reasonable implied He WD formation rates calculated with the thick H envelope models and the inclusion of these stars, suggests that at least some of the bright blue gap objects are correctly identified as He WDs.   Many of the bright blue gap objects are very bright and therefore appear to be quite young He WDs, some having cooling ages of less than 100 Myr which may indicate that there has been an increase in the production rate of He WDs over the last several Myrs as a result of the dynamical evolution of the cluster, but is still consistent with our interpretation of these objects as He WDs.  
 
 \subsubsection{Cooling Ages and Implications for CO WDs}
 \label{sec:coform}
 
 If we use this same idea to consider the formation rate of CO WDs we find that 9 of the of stars we considered CO WD candidates have cooling ages of less than 30 Myr, implying a formation rate of about 300 Gyr$^{-1}$.  This is a very small formation rate, as it is almost 2 times smaller than our lowest estimate for the turnoff rate for stars from the MS and more than 4 times smaller than the turnoff rate calculated using the HB.  However, there are also 10 ambiguous WDs of which 6 appear to fall approximately in the same range for the cooling age.  If these are added to the sample, we can calculate an implied formation rate of 500 Gyr$^{-1}$.  This is a more reasonable formation rate, but using the turnoff rate calculated from the HB stars this still seems to be rather small.  It is unclear why this is the case, but as mentioned previously, there is likely some confusion in classifying CO WDs vs. He WDs that may account for the discrepancies in the formation rates.

\subsection{Nature of the Gap Objects}
\label{sec:discgap}

\begin{figure}[htbp]
\centering
\includegraphics[width=1\textwidth]{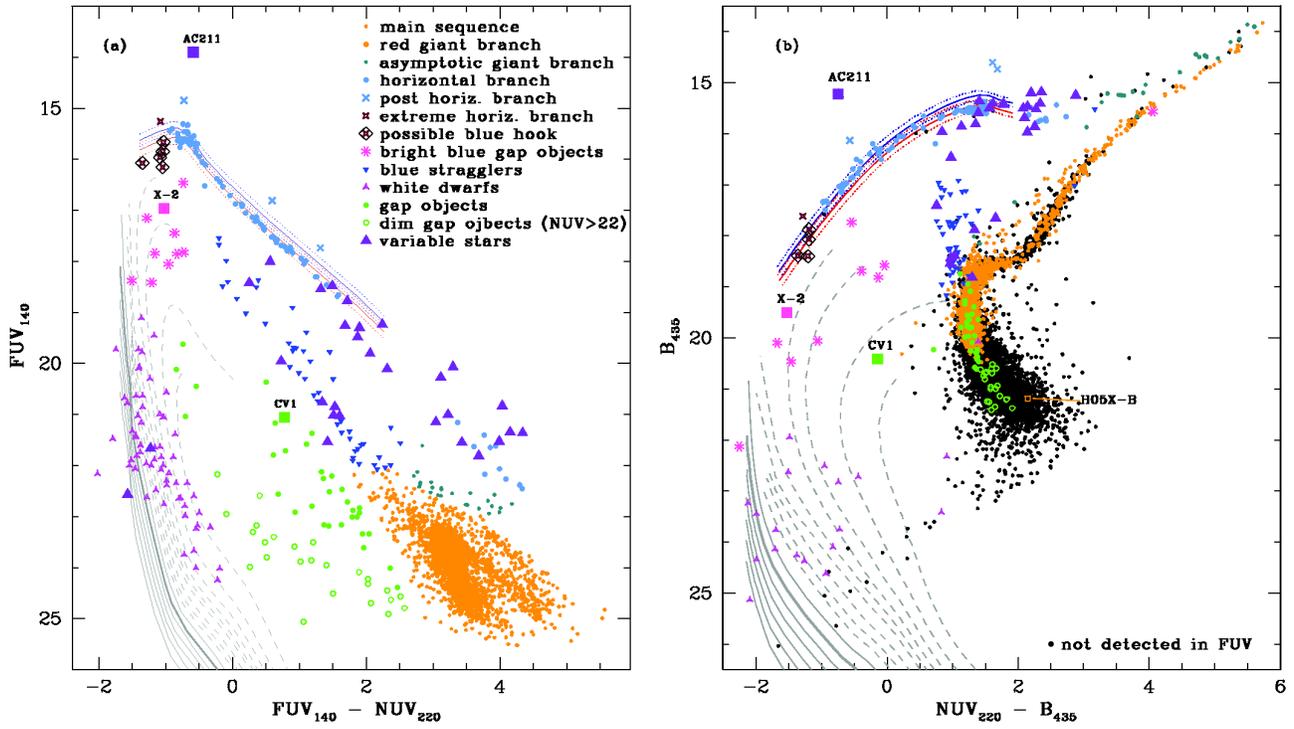}
\caption{Same as Figure \ref{fig:b_cmds} except populations distinguished and color-coded according to their positions on the FUV$_{140} - $NUV$_{220}$ (3a) CMD. See key in upper righthand corner of 3a for symbol explanation. The RGB has not been distinguished from the MS in these figures as it is very difficult to the distinguish the RGB in 3a, as discussed in the text.}
\label{fig:f_cmds}
\end{figure}

\begin{figure}[htbp]
\centering
\includegraphics[width=1\textwidth]{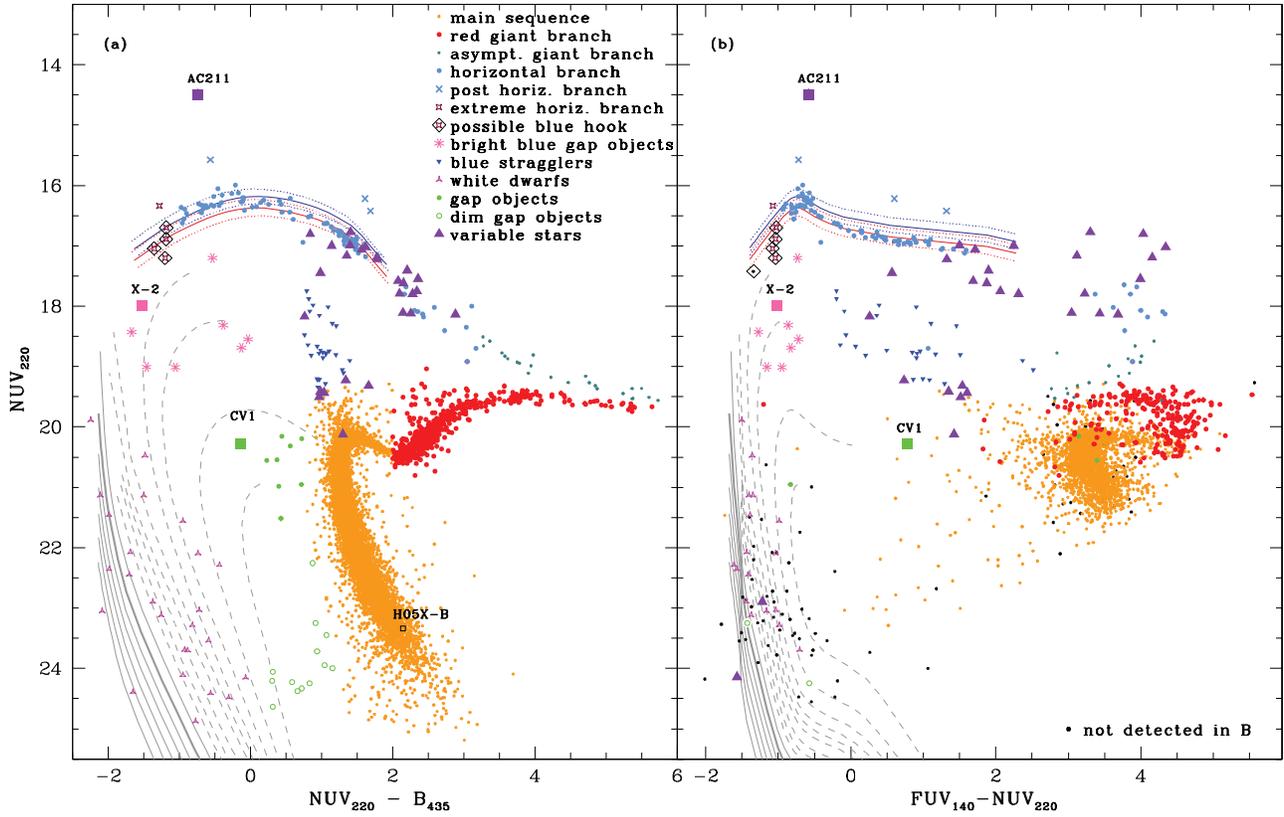}
\caption{Similar to Figures \ref{fig:b_cmds} \& \ref{fig:f_cmds} but with the magnitude axis being NUV$_{220}$ for both CMDs.  Populations distinguished and color-coded according to their position in Figure \ref{fig:b_cmds}a. See key in upper righthand corner of 4a for symbol explanations.}
\label{fig:n_cmds}
\end{figure}

\begin{figure}[htbp]
\centering
\includegraphics[width=1\textwidth]{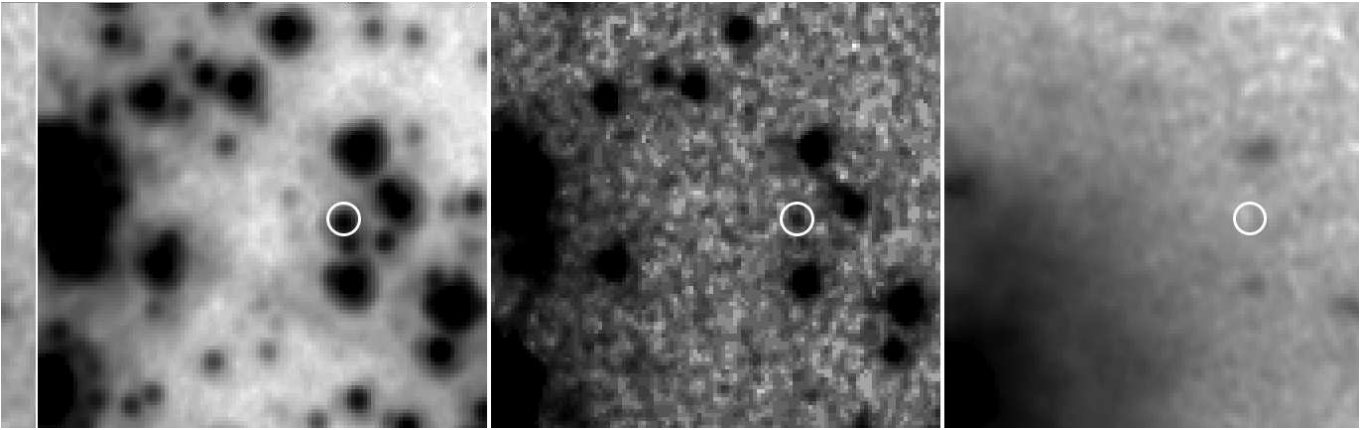}
\caption{The location of the suspected counterpart to H05X-B in B$_{435}$, NUV$_{220}$, and FUV$_{140}$ respectively.}
\label{fig:H05X-B}
\end{figure}

\begin{figure}[htbp]
\centering
\includegraphics[width=1\textwidth]{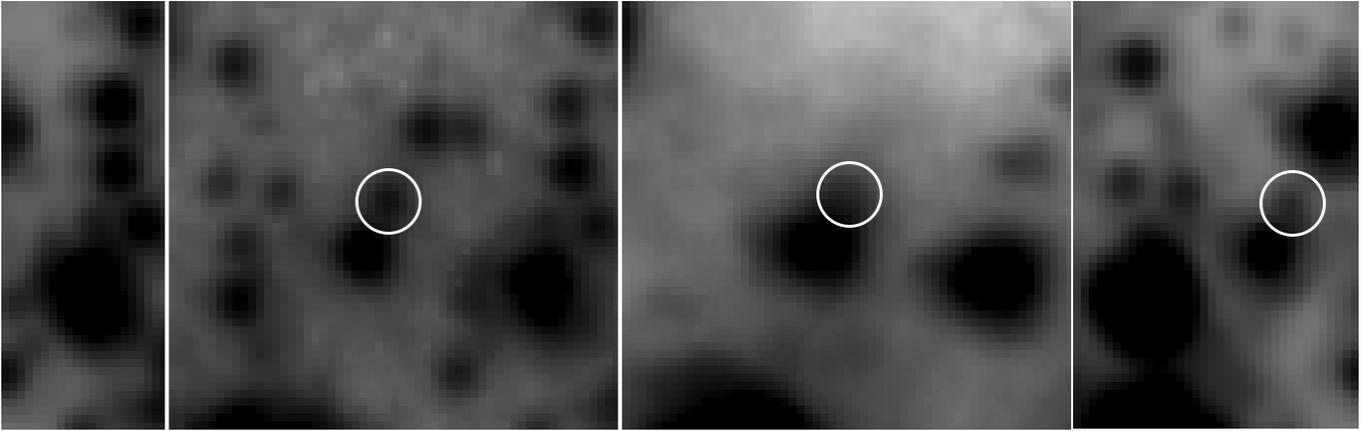}
\caption{The location of the CV1 in B$_{435}$, NUV$_{220}$, and FUV$_{140}$ respectively.}
\label{fig:CV1}
\end{figure}

In Figure \ref{fig:f_cmds}a we have identified 60 gap objects that represent possible CV candidates (bright green dots and open circles).  While this seems to be a promising set of candidates, when the B$_{435}$ photometry is added and those same stars are plotted on Figure \ref{fig:f_cmds}b, most no longer appear to be gap objects at all but instead lie on the MS.  Of the 60 gap objects identified in Figure \ref{fig:f_cmds}a, 54 were detected in the B$_{435}$ frame as well and therefore are plotted on Figure \ref{fig:f_cmds}b. Two of these stars (including CV1) appear to be gap objects in both Figures \ref{fig:f_cmds}a \& b.  The photometry for CV1 puts it clearly in the gap region on both CMDs as expected. The other star mutually identified as a gap object, hereafter UV16 (see Table \ref{tab:uv} and Fig. \ref{fig:cb}), lies very near the blue edge of the MS in Figure \ref{fig:f_cmds}b and almost as blue as the WD sequence in Figure \ref{fig:f_cmds}a.  While we can make no conclusive statement about the nature of UV16 based on photometry alone, we present it as a likely CV candidate.  

Excluding CV1 and UV16, there remain 52 gap objects that were detected in all 3 filters. Twenty-two of these have NUV$_{220}$ $>$ 22 placing them in the regime where our NUV$_{220}$ photometry is less reliable; the remaining 30 stars with reliable photometry appear as gap objects in Figure \ref{fig:f_cmds}a yet lie securely on the MS in Figure \ref{fig:f_cmds}b. While this result is puzzling, our FUV$_{140}$ and NUV$_{220}$ photometry is consistent with the similar results published by D07. In addition, these sources were inspected by eye to rule out spurious matches and false detections. Some possible explanations are discussed in the following sections.

\subsubsection{Cataclysmic Variable Candidates}
\label{sec:gapcv}

\begin{figure}[htbp]
\centering
\includegraphics[width=1\textwidth]{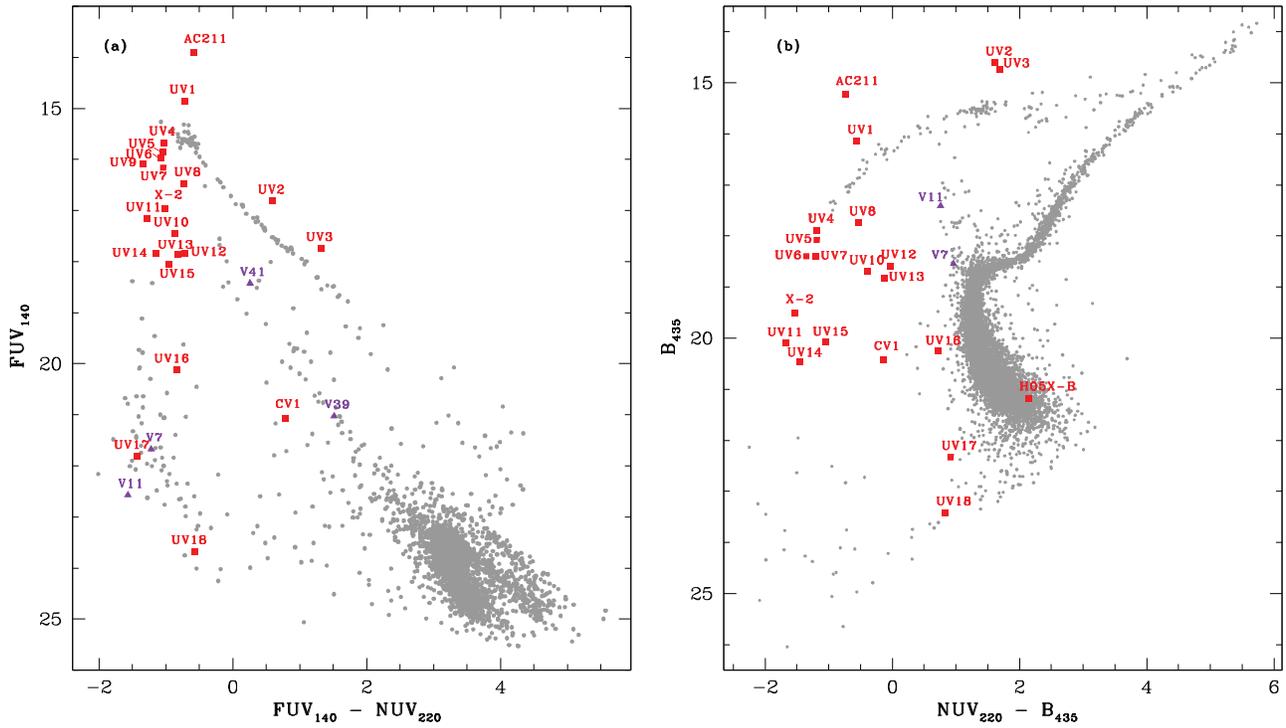}
\caption{CMD showing the stars from Tables \ref{tab:known} \& \ref{tab:uv} (red squares) as well as CV candidate variable stars from D07 (purple triangles).}
\label{fig:cb}
\end{figure}

\begin{figure}[htbp]
\centering
\includegraphics[width=1\textwidth]{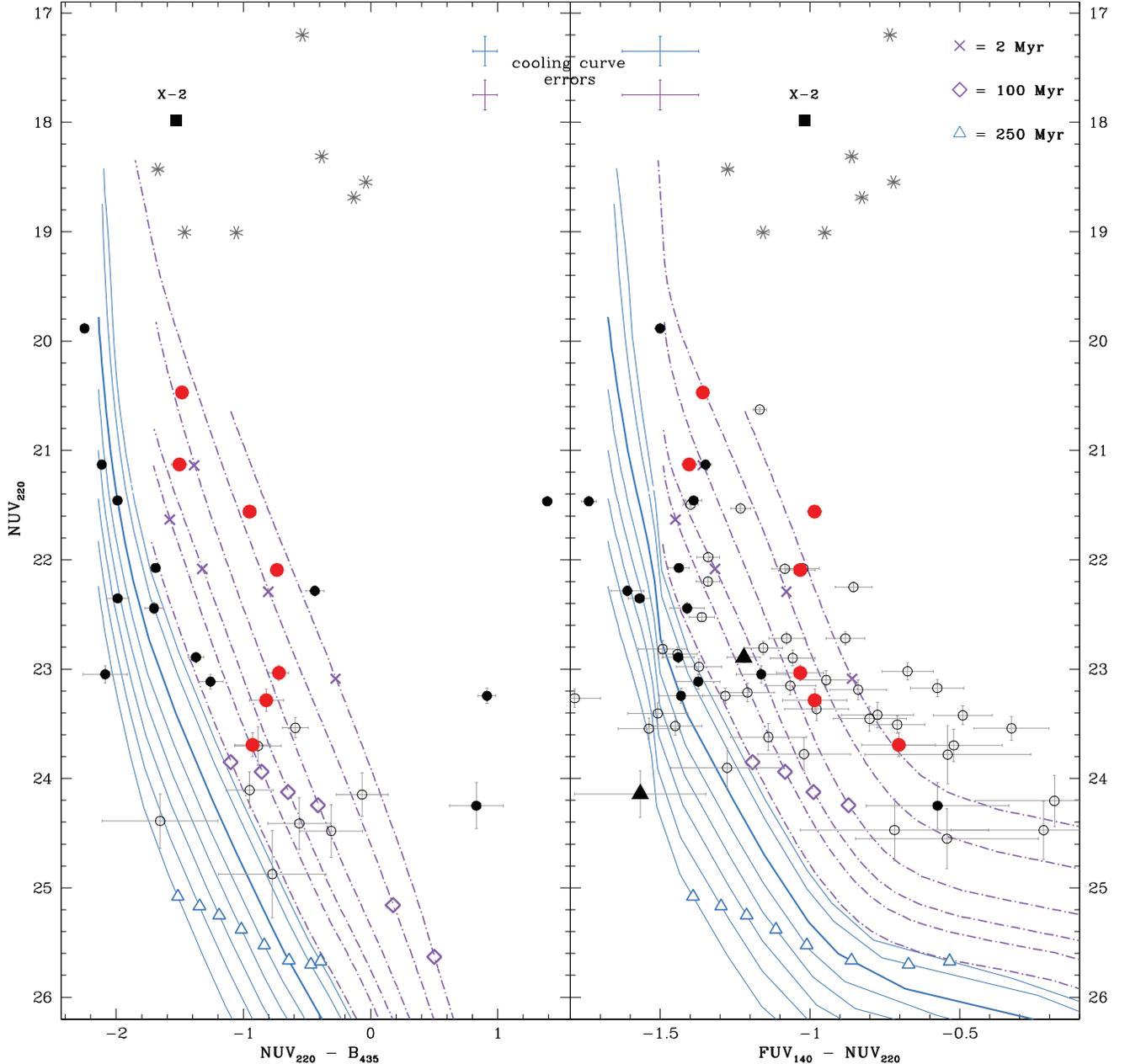}
\caption{WD candidates and model cooling sequences for CO WDs (blue solid lines) and thin H envelope He WDs (purple dot-dash lines). Masses (in M$_{\odot}$) for models, from left to right -- CO WDs: 1.10, 1.00, 0.90, 0.80, 0.70, 0.60, 0.50, 0.45; thin H envelope He WDs: 0.45, 0.35, 0.30, 0.25, 0.20, 0.175. Cooling ages are marked along the cooling curves and indicated in the key located in the upper right. The cooling curve for the fiducial 0.6 M$_{\odot}$ CO WD has been plotted as a thicker line for orientation purposes.  Filled circles are WD candidates that were detected in all three frames; open circles were only detected in two frames; larger red filled circles were detected in all three frames and represent the 7 \textit{strongest} He WD candidates (see \S \ref{sec:dischewd}); and filled triangles are stars identified as variables in D07.  The grey asterisks are bright blue gap objects that we consider possible He WD candidates based on the curves in Figure \ref{fig:wd_thick} (\S \ref{sec:dischewd} \& \ref{sec:discbrightgap}). Error bars shown were calculated by the program ALLSTAR in DAOPHOT II (see Stetson \textit{et al.} 1990). The error bars in the upper portion of the figure show the 1$\sigma$ error for the cooling curves due to the uncertainty in distance and reddening.}
\label{fig:wd_thin}
\end{figure}

\begin{figure}[htbp]
\centering
\includegraphics[width=1\textwidth]{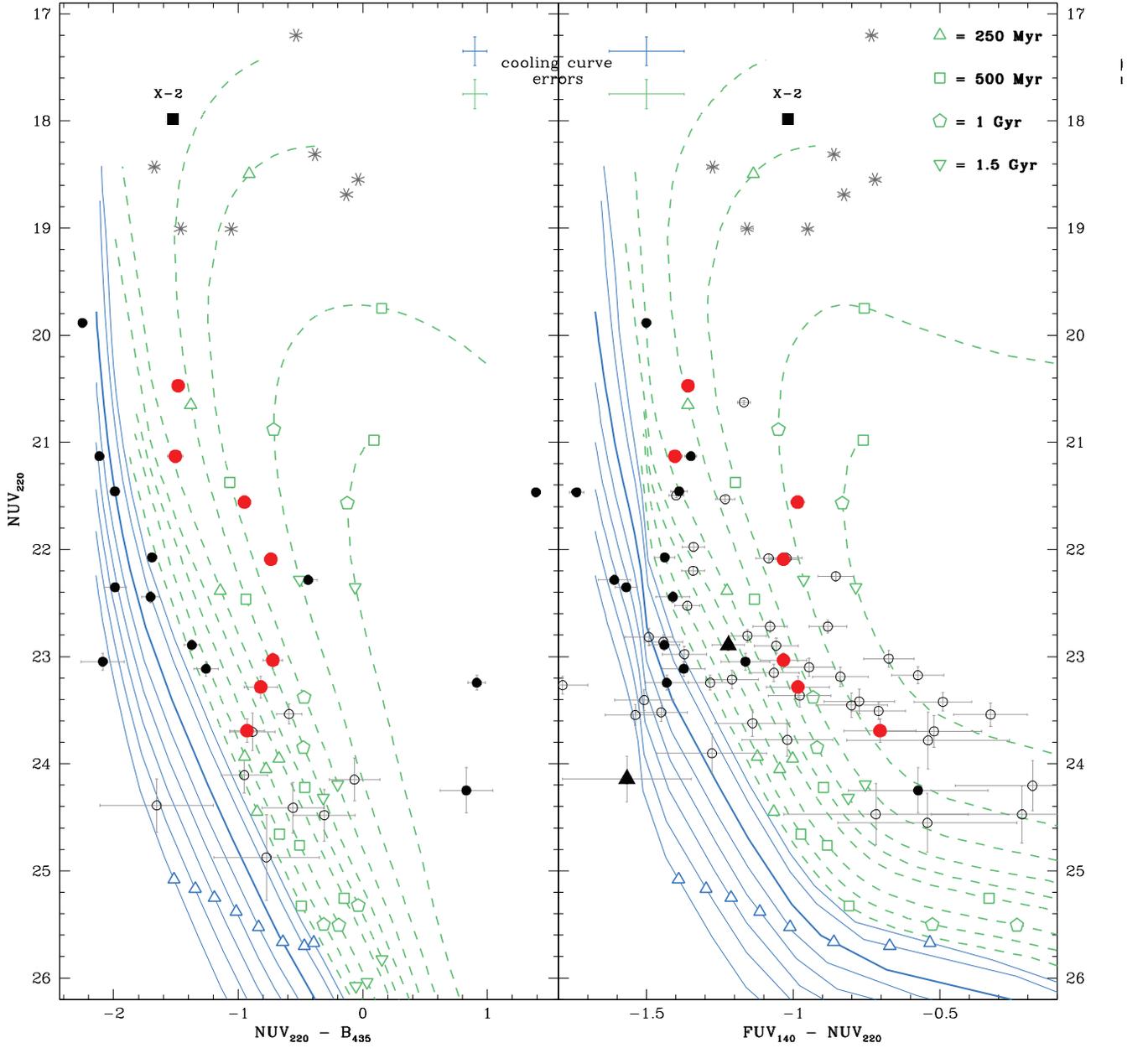}
\caption{WD candidates and model cooling sequences for CO WDs (blue solid lines) and thick H envelope He WDs (green dashed lines). Symbols for the WD candidates, bright blue gap objects, and CO WD curves are the same as in Fig. \ref{fig:wd_thin}. Masses (in M$_{\odot}$) for models from left to right -- CO WDs: 1.10, 1.00, 0.90, 0.80, 0.70, 0.60, 0.50, 0.45; thick H envelope He WDs: 0.45, 0.40, 0.35, 0.30, 0.275, 0.25, 0.225, 0.20, 0.175. Cooling ages for thick envelope He WD curves are indicated in the key located in the upper right.}
\label{fig:wd_thick}
\end{figure}

\begin{figure}[htbp]
\centering
\includegraphics[width=1\textwidth]{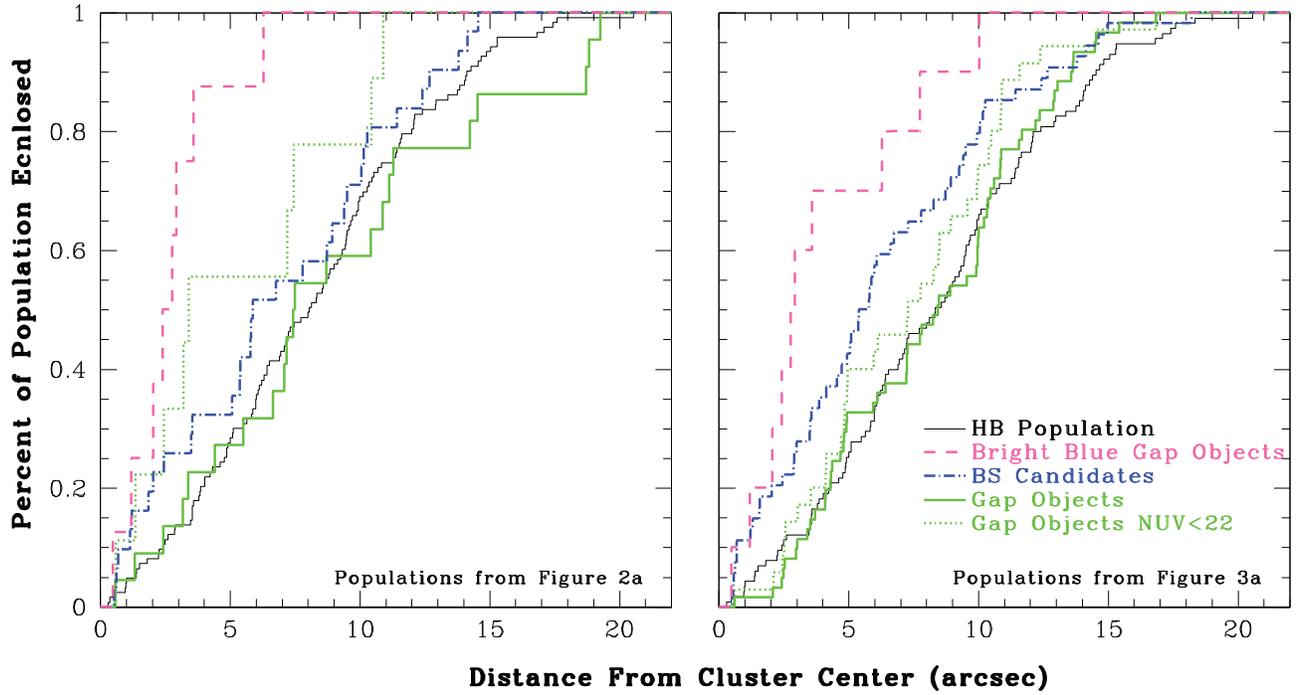}
\caption{Cumulative radial distributions for selected stellar populations.  The populations used for the left panel were selected from Fig. 2a and the populations used for the right panel were selected from Fig. 3a  (i.e.``BS Candidates" on left panel represents the distribution of those plotted as inverted blue triangles on Fig. 2 and ``BS candidates" on the right panel represents the distribution of those plotted as inverted blue triangles in Fig. 3). See key in lower righthand corner of right panel. The distribution for "Gap Objects" includes \textit{all} objects identified as gap objects (no magnitude limit).}
\label{fig:raddist}
\end{figure}

Since it is unclear whether the gap objects from Figure \ref{fig:f_cmds} are plausible CV candidates (see following sections for further discussion), we will focus our discussion of likely CV candidates to stars that appear in the gap in Figure \ref{fig:b_cmds}a.  This includes 22 objects only 6 of which were detected in FUV$_{140}$ and therefore appear on Figure \ref{fig:b_cmds}b.  These 6 include CV1 and the likely CV candidate UV16 as well 4 stars that lie either on the WD cooling sequences or MS in Figure \ref{fig:b_cmds}b. We do not consider the two that appear on the MS as probable CVs because they show no UV excess in the FUV.  The remaining two gap objects that appear on the WD cooling sequences represent much more probable CV candidates and have been included as such in Table \ref{tab:uv} (UV17 \& UV18) as well as shown on Figure \ref{fig:cb}. While both stars are dim (NUV$_{220}$ $>$ 22), they appear to have a UV excess in both NUV$_{220} -$ B$_{435}$ and FUV$_{140} -$ NUV$_{220}$ indicating the possible presence of an accretion disk.

There are also 5 somewhat luminous gap objects (21.5 $\gtrsim$ NUV$_{220} \gtrsim$ 20; 21$\gtrsim$ B$_{435} \gtrsim$ 19.5 - see Fig. 2a \& Fig. 4a) that are surprisingly not detected in FUV$_{140}$.  We individually inspected these sources in the FUV$_{140}$ frame and determined that the stars in question should be bright enough to be detected but are near bright stars and consequently lost in the extended PSF haloes.  Hence these stars also remain plausible CV candidates despite the fact that they were not detected in our FUV$_{140}$ data. 

D07 discuss several other CV candidates which they identified from light curves; these stars are identified in their paper as V7, V11, V15, V39, V40, and V41.  Our photometry produced very similar results to the ones presented in their paper although as noted previously there is a systematic magnitude difference in FUV$_{140}$ for the variable stars.  V15 and V40 were detected in FUV$_{140}$ but not NUV$_{220}$ which is consistent with their results (however since they were not detected in NUV$_{220}$ they are not included on our CMDs).  We can make no further assertions as to the nature of these variable stars as we did no investigation of the variability, but we do confirm the photometric nature as identified in D07 as being potential CVs (see Fig. \ref{fig:cb} for the location of V7, V11, V39, and V41 on our CMDs).

\subsubsection{WD-MS Detached Binary Systems}
\label{sec:gapwdms}

It is possible that at least some portion of the gap object population in Figure \ref{fig:f_cmds}a could be explained as WD-MS detached binary systems which, based on theoretical grounds alone, should be found in substantial numbers in GCs \citep[see][]{iva06}.  We performed a very basic test to explore the plausibility that the photometric properties of these gap objects could be explained as WD-MS detached binary systems.  We fit a polynomial to the MS data in each CMD using a least squares fit.  We then added the fluxes of these model MS stars to the fluxes from the model WD stars along the 0.6 M$_{\odot}$ CO WD cooling sequence to construct a set of ``observed fluxesÓ for WD-MS binaries. 

\begin{figure}[htbp]
\centering
\includegraphics[width=1\textwidth]{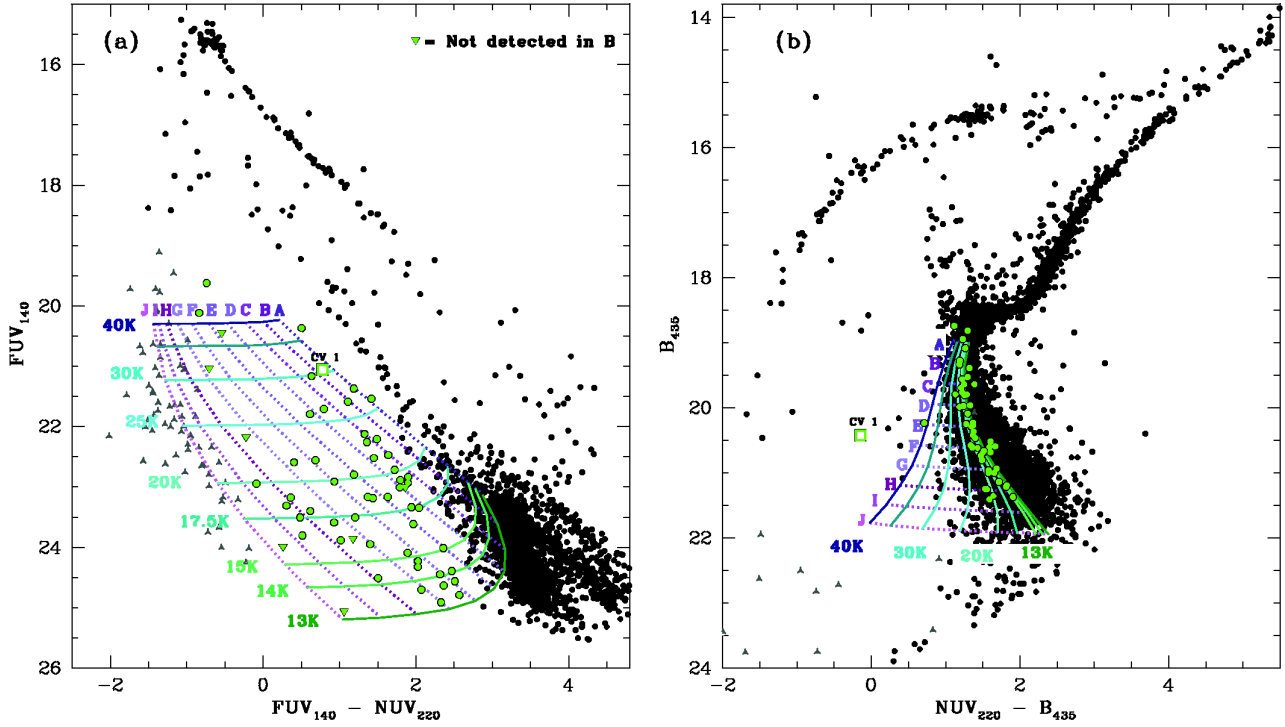}
\caption{Gap objects from Figure \ref{fig:f_cmds}a (green dots) are plotted with grids from our MS-WD detached binary system models.  Gap objects that were not detected in B$_{435}$ are plotted as green inverted triangles and stars classified as WDs are plotted as grey pinched triangles. Intersections are labeled by the effective temperature of the WD and a letter representing the MS model (see \S \ref{sec:gapwdms}); each intersection represents the resultant flux for the combination of the MS and a 0.6 M$_{\odot}$ CO WD star.  MS models range from NUV$_{220}$ $\approx$ 20.3 - 24.5 (A-J).}
\label{fig:grid}
\end{figure}

\begin{figure}[htbp]
\centering
\includegraphics[width=1\textwidth]{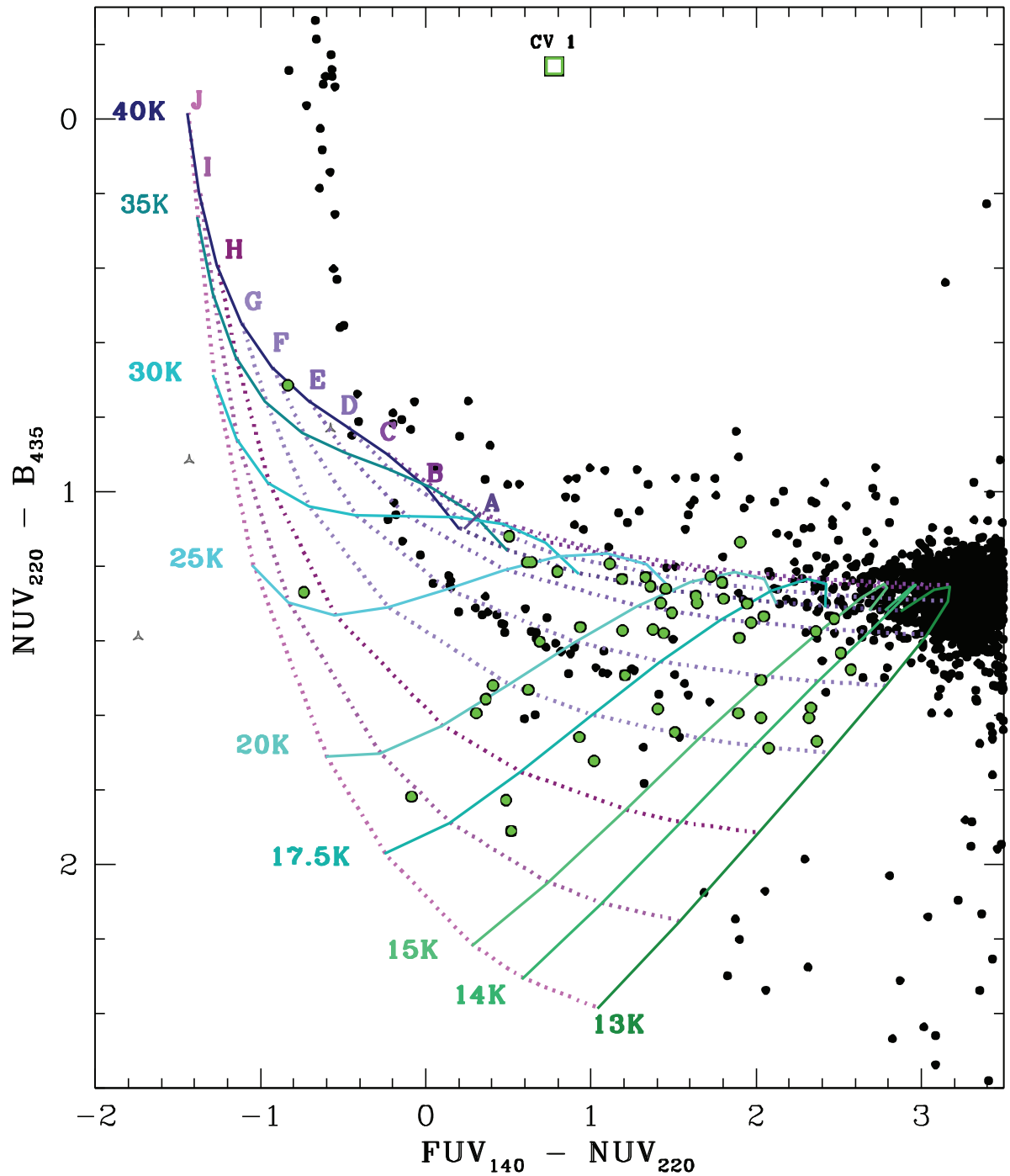}
\caption{Color-color diagram for stars detected in all three filters; symbols as in Fig. \ref{fig:grid}. For reference, MS stars appear as a clump on the right hand side, HB stars appear at the top above the model WD-MS binaries, and the BSs form a sequence near the line corresponding to WD-MS binaries that include the MS model labeled F.}
\label{fig:gridcc}
\end{figure}

In Figures \ref{fig:grid} \& \ref{fig:gridcc} we have plotted a grid where each line represents a specific model MS or WD star and each intersection represents the resulting flux from the MS $+$ WD combination.  The WD lines are solid lines labeled with the appropriate temperature for the model WD and the MS lines are dotted lines labeled with a letter so that they can easily be followed between the two diagrams.  For these models we only considered WDs between 13,000 - 40,000 K and MS stars with NUV$_{220}$ magnitudes ranging from approximately 20.3 - 24.5 (models A-J).  

From our rudimentary model in Figures \ref{fig:grid} \& \ref{fig:gridcc}, it is plausible that the observed magnitudes of the numerous gap objects from Figure \ref{fig:f_cmds}a could represent a population WD-MS detached binaries. By examining the range of MS models and WD temperatures that are populated by gap objects in Figures \ref{fig:grid}a and comparing them with Figures \ref{fig:grid}b, it can be seen that similar, though not identical, areas of our MS-WD grid are populated in both CMDs. Furthermore, the majority of these objects were examined for variability by D07 and were not found to be variable.  We therefore present this detached WD-MS binary scenario as a viable explanation for some of the gap population seen in Figure \ref{fig:f_cmds}a that seems to be absent in Figure \ref{fig:f_cmds}b.

We must point out that our grid of WD-MS detached binary models does overlap significantly into the WD region, specifically in the region where many of our He WD candidates appear (see Figure \ref{fig:grid}a).  However, we see that these same models do \textit{not} significantly overlap the WD region when examining Figure \ref{fig:grid}b, so therefore the addition of the optical filter allows us to more clearly distinguish between detached binary systems and He WD candidates.  In addition, this overlap region in Figure \ref{fig:grid}a contains a relatively large number of stars (20-25), but is only covered by models requiring the combination of WD with a very low-mass companion.  So, if our He WD candidates are in fact interpreted as WD-MS detached binaries it would indicate a strong preference for WDs to have low-mass companions; we find no obvious reason for such a preference to exist so we consider the likelihood that a significant amount of the He WD candidates are actually detached binaries to be very small. 

\subsubsection{Magnetic CVs}
\label{sec:gapamher}

It is also possible that some of gap objects in Figure \ref{fig:f_cmds}a could be magnetic CVs.  Magnetic CVs come primarily in two types: polars (AM Her stars) which have no accretion disks and intermediate polars (IPs; DQ Her stars) which have truncated accretion disks due to most, or all, of the accretion in these CV systems being directed along magnetic field lines onto the magnetic poles of the primary.  In the field the CV population is comprised of about 25\% magnetic CVs \citep{wicfer00} and it has been proposed on an observational basis that GCs contain a higher fraction of magnetic CVs than the field \citep[e.g.,][]{gri95, edm99}.  In addition, \citet{iva06} modeled the dynamical processes that lead to the formation of CVs in clusters and found that cluster CVs tend to have more massive WD primaries.  Since high mass WDs have been shown to be more likely to have strong magnetic fields \citep[e.g.,][and references therein]{lie88, sch03} this may explain the magnetic nature of clusters CVs.  

Since the accretion disk contributes strongly to the flux of non-magnetic CVs, polars and IPs will have different photometric properties than non-magnetic CVs such as dwarf novae (like CV1) and we argue that the missing accretion disk could account for the lack of NUV excess in Figure \ref{fig:f_cmds}b.  In FUV$_{140}$ the WD dominates the flux while the MS companion dominates the flux in B$_{435}$; the lack of an accretion disk leads to a significant decrease in the FUV$_{140}$ and NUV$_{220}$ flux compared to that of an accretion disk system and would cause the system to have colors more similar to that of a WD-MS detached binary.  \citet{edm99} find that one of the CVs identified as a magnetic CV in their paper appears on the MS in the V vs. V$-$I CMD. Additionally, in recent results from GALEX observations of open cluster M67, EU Cnc, an AM Her system \citep{nair05}, has been identified to have very similar photometric properties to the perplexing gap objects identified here (Hamper et al. 2010, in preparation).  

The literature contains ideas invoking the magnetic nature of globular cluster CVs to explain observed differences between field CVs and cluster CVs, such as the relatively red optical colors and high optical to X-ray flux ratios \citep[e.g.,][and references therein]{edm99, dob06}. \citet{edm03} and \citet{dob06} also propose that a combination of low accretion rates and strong magnetic moments of WDs in magnetic CVs may stabilize the disk such that outbursts would be rare, possibly explaining the low rate of observed outbursts in cluster CVs \citep[e.g.,][and references therin]{sha96, dob06}.  We therefore suggest that it is quite possible that some of our gap objects do represent a population of magnetic CVs. 

X-ray luminosity measurements would be particularly advantageous in constraining the nature of these objects.  Currently 6 sets of observations of M15 from the Chandra X-ray Observatory are available in the archives. For a complete list of the available observations and limiting X-ray luminosities associated with each, the reader is referred to \citet{hei09}.  Despite the generous amount of X-ray data available for M15, the region of most interest for this paper, the central 20\arcsec $\times$ 20\arcsec, is mostly dominated by two extremely bright sources (AC211 and X-2) that dominate most of the dimmer sources in this region -- see Fig. 1 of  \citet{hei09}.

\subsubsection{Dim X-ray Source H05X-B}
\label{sec:H05XB}

\citet{han05} used archival HST images (from the Wide Field/Planetary Camera 2) to identify a possible counterpart to the suspected DN H05X-B (see Fig. 6 in that paper).  The source that we have identified as H05X-B is consistent with the position of the suggested counterpart for ``source B" in their paper.  However, the color of this star does not appear to have any blue excess (see Fig. \ref{fig:b_cmds}-\ref{fig:n_cmds} and Fig. \ref{fig:cb} of this paper). \citet{han05} identify H05X-B as a possible DN or qSXT, yet the color that we measure (NUV$_{220} -$ B$_{435}$ = 2.15) is redder than expected for DNe or SXTs, even those observed during quiesence (qSXT: Shahbaz et al. 2003; Edmonds et al. 2002; DNe: Bailey 1980).  It is possible that the star that we have chosen is not the true counterpart. Several other sources are within the error circle shown in \citet{han05}, but we examined each of these and none seem to have particularly blue colors either.

\citet{han05} report this star as blue in their U-V color, so we do not rule out that the source identified as H05X-B could be the true counterpart.  It is possible that the NUV$_{220}$ observations in our data set happened to coincide with a particularly quiescent state.  However, since the reported colors in the literature for quiescent DNe are bluer than those for qSXTs, it seems more plausible that H05X-B may be a qSXT, which is contrary to the conclusion made by \citet{han05}.

\subsection{Post-HB Stars}
\label{sec:discpagb}
We have identified 3 stars, aside from AC211, that lie at least 0.5 mag above the horizontal branch (see light blue X's in Fig. 2-4 or UV1-3 in Table 2 and Fig. 7). We tentatively identify these stars as being in post-HB evolutionary states and P-EAGB or AGB-manqu\'{e} candidates.  As these stars are only between 0.5 and 1 mag brighter than the HB, it is possible they have not yet exhausted core helium and are just beginning to evolve off the HB.  Spectra for these stars or careful comparison with post-HB models is necessary to confirm the nature of these stars, but is beyond the scope of this paper.  Judging solely on their relative location on the CMDs in Figure \ref{fig:n_cmds}, we present UV1 as an AGB-manqu\'{e} candidate and UV2 and UV3 as P-EAGB candidates  as these stars appear roughly consistent with the [Fe/H] = $-2.3$ models of \citet{dor93}.  

Since P-EAGB and AGB-manqu\'{e} stars are considered the progeny of the EHB, this raises the question of whether we should expect to see any such stars in M15 because we have a very small sample of EHB candidates.  For this paper we defined the EHB as stars with T$\rm_{eff}$ $\geq$ 20,000K, but in the low-metallicity models of \citet{dor93} some stars with T$\rm_{eff}$ as low as 10,000K do not reach the thermal pulsation phase of the AGB and therefore become P-EAGB stars.  \citet{dor93} find the ratio of the post-HB lifetime to HB lifetime for AGB-manqu\'{e} stars to be between 1:5 and 1:6. For P-EAGB stars this ratio varies considerably depending on luminosity, with the dimmest  having the longest post-HB lifetime, but the ratio of lifetimes is usually about an order of magnitude less than that of AGB-manqu\'{e} stars \citep[see also][]{ber95}.  

If we treat our entire population of EHB candidates as true EHB stars (instead of BHks), then we would expect to find 1 AGB-manqu\'{e} star.  However, it would seem highly unlikely to also find 2 P-EAGB stars. However, if we include stars with  T$\rm_{eff}$ $\geq$ 15,000K as possible progenitors of P-EAGB stars, we have 16 candidate progenitors.  And following \citet{dor93} decreasing our temperature to  T$\rm_{eff}$ $\geq$ 10,000K, we end up with 37 candidate P-EAGB progenitors.  Therefore, we do not rule out these stars as P-EAGB candidates, but we caution that it is unclear whether HB stars as cool as T$\rm_{eff}$ = 10,000K or 15,000K  could produce the P-EAGB candidates identified here.  

\subsection{Nature of Bright Blue Gap Objects}
\label{sec:discbrightgap}

The nature of the objects identified as bright blue gap objects is very hard to determine from photometric properties alone as they do not obviously belong to any standard population and they populate a region of the CMD in which stars following canonical stellar evolution do not exist for any substantial period of time.  From Figure \ref{fig:b_cmds}b, the 7 stars, excluding X-2, that were detected in all three filters and appear as bright blue gap objects on all CMDs (UV8 \& UV10-15 in Table 2 and Fig. 7) appears as though they could feasibly be members of the HB or BS sequence; however, Figure \ref{fig:b_cmds}a illustrates that this is highly unlikely.  The color of these objects makes them possible CV candidates, but this also seems an unlikely explanation because 4 of the 7 stars are brighter than the MSTO and several magnitudes brighter than CV1 in all three filters.  The remaining 3 have a B$_{435}$ magnitude similar to CV1 and the MSTO, but are significantly brighter in NUV$_{220}$ and FUV$_{140}$. 

The radial distribution of the bright blue gap objects (Fig. \ref{fig:raddist}) indicates that these stars are centrally concentrated and therefore are likely massive binary systems.  This, combined with the fact that X-2 (an ultra-compact LMXB)  lies in a similar region of the CMD, leads us to consider that they may be close binary systems with some current mass transfer.  If this is the case and the population \textit{is} related to accretion disk phenomena it is expected that at least some of these stars should be variable.  Although these stars were all included in D07, none were identified as variables.  While this is not favorable for the accretion disk hypothesis, X-2 was not selected as variable in D07 either, as the amplitude of its variability is less than 0.2 mag and therefore too small to be selected as a variable in their study. We, therefore, can not rule out these objects as potential mass-transfer binaries but find this explanation somewhat weak because none of these stars were found to show strong FUV$_{140}$ variability and none have yet been identified as sources of X-ray emission.

The dimmest three bright blue gap objects (UV13, UV14, \& UV15; Table \ref{tab:uv}) appear to possibly be associated with the WD sequence, but are more than 1.5 magnitudes brighter than the other white dwarfs in B$_{435}$ making this seem unlikely as well.  Based on their position in the CMD alone, one could infer that the entire bright blue gap object population could be related to the BHk stars.  But again, we rule this out as a likely explanation because BHk stars are not expected to be found more than approximately 1 magnitude dimmer than the ZAHB in UV filters \citep[][ and references therein]{bro01};  it can be seen in all the CMDs that the bright blue gap objects are significantly more than 1 mag. dimmer than the equivalent temperature BHB stars in all three filters. UV8 seems to be the most likely BHk candidate of the bright blue gap objects judging by its location in the FUV$_{140} -$ NUV$_{220}$ CMDs, however its location in the NUV$_{220} -$ B$_{435}$ diagrams makes its nature unclear.  

Finally, we return to the possibility that some of the bright blue gap objects might be young He WDs.   As is most apparent in Figure \ref{fig:wd_thick} these objects appear to be roughly consistent with the early stages of the He WD cooling sequence for 0.200$-$0.275 M$_{\odot}$ He WDs in the thick H envelope models.  In our analysis in \S \ref{sec:dischewd} we included these stars as plausible He WD candidates when analyzing the cooling ages and implied formation rates and found that while the cooling ages implied for these stars require a somewhat increased production rate of He WDs over the last several 100 Myrs, this interpretation seems very reasonable. 

We find it most likely that the majority of the bright blue gap objects are either young He WDs or somehow related to mass transfer binaries with current accretion disk phenomena.  We find the former to be the more convincing explanation but do not rule out that the bright gap population may encompass two or more physically distinct populations, especially since it does include X-2. 

\section{DISCUSSION}
\label{sec:discussion}
 
\subsection{Gap Population}
The filter combination of FUV$_{140}$ and NUV$_{220}$ was very successful in identifying members of the gap zone, which are potential CVs as pointed out by D07.  However, the addition of the B$_{435}$ filter has made the nature of these stars less clear.  Because the majority of these stars appear on the MS in NUV$_{220} -$ B$_{435}$ CMDs, it seems unlikely that they are standard DNe CVs.  We find the most likely explanation to be that this group is composed of a combination of magnetic CVs and WD-MS detached binaries.  There is no analog in the literature for such a significant population of FUV-bright objects that appear MS-like in a color such as NUV$_{220} - $B$_{435}$; so we can draw only weak conclusions about the gap population we have identified in Figure \ref{fig:f_cmds}a without the addition of H$\alpha$ imaging to seek out emission line sources or spectroscopic follow up.

\subsection{M15 as a Blue Hook Cluster}
\label{sec:m15bhk}
We have identified 5 of the EHB stars as potential BHk stars.  If we are correct in our classification of these stars as BHk stars, it may provide an important data point in understanding the properties of BHk clusters. \citet{die09} investigated trends among clusters containing BHk stars with their strongest result being that BHk stars seem to exist exclusively in the most massive clusters. They note that this might be attributed to a natural bias toward massive clusters, since intrinsically rare stars are necessarily more likely to be found in larger samples.  They also found weaker correlations between the presence of BHk stars and other cluster properties such as concentration parameter, core radius, and relaxation time, but considered the most significant correlation to be that with cluster mass.  M15 was included in their sample, but it was considered a non-BHk cluster as it was only known to have one BHk star in the outer region (see \S \ref{sec:bhk}). Had M15 been included as a BHk cluster in that study it would have had a strong lever arm on the results, because, of the clusters in their study that had 4 or more BHk stars, M15 has the lowest mass, highest central density, shortest core relaxation time, smallest core radius, and highest concentration parameter.  However, it is unclear whether M15 would be considered a BHk cluster even with all the candidates we have identified here included since all the ``BHk clusters" in the \citet{die09} sample had 10 or more BHk stars. Yet, we have only searched a small region in M15, so there may be more BHk stars that we have not identified. Nevertheless, irrespective of how we define a BHk cluster, it is clear that a BHk population in M15 provides an important constraint in understanding the origin of BHk stars.

Since we feel that the existence of a BHk population most likely weighs in favor of the late He-flash scenario, as discussed in \S \ref{sec:discbhk}, it should be noted that within this scenario the expected mass range for BHk stars is quite small.  \citet{milber08} calculate that the expected mass of a low-metallicity remnant capable of becoming a BHk star via a late He-flash lies in a very narrow range between 0.48 - 0.50 M$_{\odot}$ (see that paper for more details).  It is expected that there would be a spread in the total mass loss from the progenitors of these BHk stars thus resulting in a comparable number of stars with post-MS ages similar to BHk stars ($\lesssim$ 100 Myr) that just ``miss" becoming BHk stars and end up as either massive He WDs (which just avoid He-core ignition) or low-mass CO WDs (which arrive as such following a phase as an EHB or BHk star).  It is unclear whether M15 contains such a population. There are a significant number of potential WDs that may be either low-mass CO WDs or massive He WDs in the appropriate age range (see Fig. \ref{fig:wd_thin} \& \ref{fig:wd_thick}), but many of these candidates have ended up classified as ``Ambiguous WDs" (Table \ref{tab:wd}) or were only detected in two frames due to the intrinsic photometric uncertainty associated with such dim stars.  So, without a more precise determination of both the mass and age of our WD candidates, we can make no further claim to the existence of such a population other than to note its expected presence.  

\section{SUMMARY \& CONCLUSIONS}
\label{sec:conclusions}
We have presented a photometric identification and analysis for several UV-bright populations in the central region of the post-core collapse globular cluster M15. We additionally have included photometry for 4 previously identified X-ray sources: M15 CV1, AC211, M15 X-2, and H05X-B.  Our work has elaborated on the work of \citet{die07}, who analyzed the FUV$_{140}$ and NUV$_{220}$ images. We reanalyzed these images and added the B$_{435}$ filter and the NUV$_{220} -$ B$_{435}$ color which has added further insight into the nature of the populations discussed here. The UV-bright populations we have analyzed include many stages of non-canonical stellar evolution including blue stragglers, extreme horizontal branch stars, blue hook stars, helium-core white dwarfs, and cataclysmic variable candidates. 

We have selected 53 blue straggler candidates which display a clear central concentration as expected for BSs. Since the expected zone of avoidance is beyond the field of view for our images we are unable to investigate whether the BS population displays the double peaked bimodal distribution that has been discussed in the literature for several other clusters. We found 60 CV candidates populating the gap between the MS and WD region, consistent with the findings of D07, however upon inclusion of the B$_{435}$ filter we found that many of these stars do not display expected CV colors, but instead appear as MS stars.  Thus, we suggest that these gap objects may be magnetic CVs with truncated or absent accretion disks or possibly detached WD-MS binaries. However, we also find three gap objects that we consider very likely CV candidates (UV16-18) as they display colors similar to what is expected for a CV.  

We have used a ZAHB model to select 6 extreme horizontal branch candidates, 5 of which appear to be subluminous in the UV and therefore better candidates for blue hook stars.  Due to the existence of these candidates, in addition to the one previously identified BHk star, we suggest that M15 be considered a BHk cluster for future studies on clusters containing BHk stars as it may provide important constraints on how these stars are formed.  We also identify three stars that represent plausible post-HB stars that may be AGB-manqu\'{e} or post-early AGB stars, the progeny of EHB stars.  Our identification of these post-HB stars however is very preliminary as their photometric properties seem consistent with AGB-manqu\'{e} and P-EAGB states, yet statistically it seems improbable to have detected three stars in these relatively short lived stages of evolution.   

Additionally, we uncovered a population of ``bright blue gap objects" for which there is no obvious analog in the literature.  We consider these stars to most likely be young He WDs, but they could be related to the BHk population or accretion disk binary systems.  UV10-UV15 seem to be the most plausible candidates for He WDs. There is no obvious population to which UV8 belongs but it seems to be related to the HB and is a possible BHk candidate.  In addition to the bright blue gap objects, we have identified a significant population of candidate He WDs which we analyzed using model He WD cooling sequences.  We have probed quite deep into the He WD sequence and find 7 strong He WD candidates detected in all three images; an additional 38 candidates were detected in only two images. This is the first strong evidence of the existence of a He WD population in M15.  We analyzed both thin and thick H envelope models and based on the cooling ages and overall fit to the population we find the thick H envelope models to be better fit to our data.  The formation rates suggested by the thin H envelope models are unreasonable unless we have misidentified several of our candidate He WDs.  The formation rates implied by the thick H envelope models range from 7-37 He WDs produced each Gyr.  The collision timescale of red giants in the core of M15, is high enough that it is possible that collisions account for a significant fraction of the He WDs.  Furthermore, the implied lower limit on the binary fraction calculated from these formation rates suggest that close binary systems are also likely to contribute to the formation of He WDs in M15.

\clearpage
\begin{center}
\begin{deluxetable}{c c c c c c}
\tabletypesize{\small}
\tablewidth{0.85\textwidth}
\tablecaption{Previously Identified X-ray Binary Systems}
\tablehead{\colhead{Star} & \colhead{R.A.} & \colhead{Dec.} & \colhead{B$_{435}$} & \colhead{NUV$_{220}$} & \colhead{FUV$_{140}$}} 
\startdata
AC211$^{a}$ & 21:29:58.323 & +12:10:01.94 & 15.23 & 14.49 & 13.91\\
X-2$^{b}$ & 21:29:58.142 & +12:10:01.52 & 19.51 & 17.98 &16.96\\		  
CV1$^{c}$ & 21:29:58.357 & +12:10:00.33 & 20.42 & 20.28 & 21.06\\
HX05-B$^{d}$ & 21:29:58.323 & +12:10:11.69 & 21.19 & 23.33 & -	
\enddata
\tablerefs{$^{a}$\citet{aur84}; $^{b}$ \citet{die05};  $^{c}$\citet{sha04}; $^{d}$\citet{han05}}
\label{tab:known}
\end{deluxetable}
\end{center}
\begin{center}
\begin{deluxetable}{c c c c c c c}
\tabletypesize{\small}
\tablewidth{1\textwidth}
\tablecaption{UV-Bright Stars of Interest}
\tablehead{\colhead{Star ID} & \colhead{x$_{F220W}$} & \colhead{y$_{F220W}$} & \colhead{B$_{435}$} & \colhead{NUV$_{220}$} & \colhead{FUV$_{140}$} & \colhead{Candidate Classification}}
\startdata
UV1 & 233.98 & 93.13 & 16.14 & 15.57 & 14.85 & AGB-manqu\'{e}\\		  
UV2 & 341.12 & 920.76 & 14.61 & 16.22 & 16.82 & P-EAGB\\		  
UV3 & 579.31 & 500.38 & 14.74 & 16.42 & 17.74 & P-EAGB\\		  
UV4 & 773.86 & 722.57 & 17.88 & 16.70 & 15.67 & BHk\\			  
UV5 & 65.31 & 1016.84 & 18.07 & 16.89 & 15.85 &BHk\\			  
UV6 & 672.66 & 595.49 & 18.40 & 17.04 & 15.97 & BHk\\			  
UV7 & 932.36 & 976.88 & 18.40 & 17.20 & 16.16 & BHk\\			  
UV8 & 688.90 & 696.32 & 17.74 & 17.20 & 16.47 & Unknown\\ 		  
UV9 & 1028.58 & 1072.23 & - & 17.42 & 16.08 & BHk\\
UV10 & 636.65 & 635.43 & 18.70 & 18.31 & 17.45 & He WD\\ 
UV11 & 548.94 & 603.81 & 20.10 & 18.43 & 17.16 & He WD\\
UV12 & 630.50 & 602.71 & 18.58 & 18.55 & 17.83 & He WD\\  
UV13 & 629.48 & 698.07 & 18.82 & 18.69 & 17.86 & He WD\\  
UV14 & 582.81 & 643.68 & 20.47 & 19.01 & 17.85 & He WD\\  
UV15 & 412.42 & 687.64 & 20.07 & 19.01 & 18.06 & He WD\\  
UV16 & 209.254 & 608.48 & 20.24 & 20.95 & 20.12 & CV\\
UV17 & 626.60* & 369.06* & 22.33 & 23.24 & 21.81 & CV\\
UV18 & 212.69* & 980.50* & 23.42 & 24.25 & 23.68 & CV
\enddata
\tablecomments{The coordinates listed as x$_{F220W}$ \& y$_{F220W}$ are x, y coordinates for the sources in the archival NUV$_{220}$ image ``j8si08010\_drz.fits. " The coordinates for UV17 \& UV18 are marked with asterisks as these sources are both very dim and therefore difficult to identify in the this frame, but are clearly present in the fully combined images we used as our master images.}
\label{tab:uv}
\end{deluxetable}
\end{center}
\begin{center}
\begin{deluxetable}{c c c c}
\tabletypesize{\small}
\tablewidth{0.8\textwidth}
\tablecaption{Summary of WD candidates}
\tablehead{\colhead{ } & \colhead{Total Candidates} & \colhead{Detected in All Three Frames}}
\startdata
\textbf{All WDs}$^{1}$ & \textbf{73} & \textbf{20} \\
CO WDs & 11 & 5 \\
He WDs & 45 & 7 \\
Ambiguous WDs$^{2}$ & 10 & 5 \\
Possible WDs$^{3}$ & 5 & 3 \\
D07 Variables & 2 & - \\
Bright He WDs$^{4}$ &  7 & 7
\enddata
\tablecomments{$^{1}$ ``All WDs" does not include the bright blue gap objects, as this population is disconnected from the canonical WD region. $^{2}$ ``Amibguous WDs" are strong WD candidates which lie near the transition between the CO and He WD cooling sequences, making their identification as a CO vs. He WD very difficult. $^{3}$ ``Possible WDs" are candidates with significant photometric error or inconsistencies such that their nature is unclear.  $^{4}$ ``Bright He WDs" are bright blue gap objects that appear consistent with He WD cooling sequences.}
\label{tab:wd}
\end{deluxetable}
\end{center}
\clearpage
\end{document}